\documentclass{article}
\usepackage[a4paper, margin=2.6cm]{geometry}
\usepackage[utf8]{inputenc}
\usepackage{amsmath, amsfonts,amsthm}
\usepackage{float}
\usepackage{cite}
\usepackage{graphicx}
\usepackage{textcomp}
\usepackage{xcolor}
\usepackage{tikz}
\usepackage{pgfplots}
\usepackage[export]{adjustbox}
\usetikzlibrary{spy}
\usepackage{graphicx}
\usepackage{tabularx}
\usepackage{xcolor}
\usepackage{amssymb,amsmath,amsfonts, amsthm}
\usepackage{bm}
\usepackage{tikz}
\usepackage{multirow}
\usepackage{booktabs}
\usepackage{blindtext}
\usepackage{multicol}
\usepackage{tikzscale}
\usepackage{numprint}
\npdecimalsign{.}
\nprounddigits{3} 
\usepackage{caption}
\usepackage{subcaption}
\usepackage{enumitem}
\usepackage{float}
\usepackage{pict2e}
\usepackage{siunitx}
\usepackage[english]{babel}
\usepackage{authblk}

\usepackage{array}   
\renewcommand{\arraystretch}{2.5}
\usepackage[most]{tcolorbox}
\definecolor{darkspringgreen}{rgb}{0.09, 0.45, 0.27}
\usepackage[frozencache=true,cachedir=minted-cache]{minted}

\usepackage{mathtools}
\usepackage[percent]{overpic}
\usepackage{algorithm,algcompatible,amsmath}

\DeclareMathOperator*{\argmin}{arg\,min}
\algnewcommand\INPUT{\item[\textbf{Input:}]}%
\algnewcommand\PARAMETER{\item[\textbf{Parameters:}]}%
\algnewcommand\OUTPUT{\item[\textbf{Output:}]}%

\usepackage[
 colorlinks=true, 
]{hyperref}

\usepackage{cleveref}

\usepackage{algorithmicx}
\usepackage{comment}
\usepackage{array}

\usepackage{algorithm}
\usepackage{algpseudocode}
\algnewcommand{\Inputs}[1]{%
  \State \textbf{Inputs:}
  \Statex \hspace*{\algorithmicindent}\parbox[t]{.8\linewidth}{\raggedright #1}
}
\algnewcommand{\Initialize}[1]{%
  \State \textbf{Initialize:}
  \Statex \hspace*{\algorithmicindent}\parbox[t]{.8\linewidth}{\raggedright #1}
}

\usepackage[most]{tcolorbox}
\definecolor{darkspringgreen}{rgb}{0.09, 0.45, 0.27}

\title{A directional regularization method for the limited-angle Helsinki
Tomography Challenge using the Core Imaging Library (CIL)}

\author[1,2]{Jakob Sauer J\o{}rgensen}
\author[2,3]{Evangelos Papoutsellis}
\author[4]{Laura Murgatroyd}
\author[4]{Gemma Fardell}
\author[4]{Edoardo Pasca}

\affil[1]{Department of Applied Mathematics and Computer Science, Technical University of Denmark. Richard Petersens Plads, Building 324, 2800 Kgs. Lyngby, Denmark.}
\affil[2]{Department of Mathematics, The University of Manchester, Oxford Road, Alan Turing Building, Manchester M13 9PL, UK}
\affil[3]{Finden Ltd, Rutherford Appleton Laboratory,  Harwell Campus, Didcot OX11 0QX, United Kingdom}
\affil[4]{Scientific Computing Department, Science \& Technology Facilities Council, Rutherford Appleton Laboratory, 
Harwell Campus,
Didcot
OX11 0QX, UK}

\newcommand{\sysmat}{\bm{A}}
\newcommand{\dmat}{\bm{D}}
\newcommand{\sino}{\bm{b}}
\newcommand{\im}{\bm{u}}
\newcommand{\zerovec}{\bm{0}}
\newcommand{\onevec}{\bm{1}}
\newcommand{\maskvec}{\bm{m}}
\newcommand{\indifun}[1]{I_{#1}}
\newcommand{\mua}{\mu_\mathrm{A}}
\newcommand{\mm}{\mathrm{mm}}

\begin{document}

\date{\vspace*{-1cm}}

\maketitle

\begin{abstract}
This article presents the algorithms developed by the Core Imaging Library (CIL) developer team for the Helsinki Tomography Challenge 2022. The challenge focused on reconstructing 2D phantom shapes from limited-angle computed tomography (CT) data. The CIL team designed and implemented five reconstruction methods using CIL (\url{https://ccpi.ac.uk/cil/}), an open-source Python package for tomographic imaging. 
The CIL team adopted a model-based reconstruction strategy, unique to this challenge with all other teams relying on deep-learning techniques.
The CIL algorithms showcased exceptional performance, with one algorithm securing the third place in the competition. The best-performing algorithm employed careful CT data pre-processing and an optimization problem with single-sided directional total variation regularization combined with isotropic total variation and tailored lower and upper bounds.
The reconstructions and segmentations achieved high quality for data with angular ranges down to 50 degrees, and in some cases acceptable performance even at 40 and 30 degrees.
This study highlights the effectiveness of model-based approaches in limited-angle tomography and emphasizes the importance of proper algorithmic design leveraging on available prior knowledge to overcome data limitations. Finally, this study highlights the flexibility of CIL for prototyping and comparison of different optimization methods.\footnote{Corresponding author: Jakob S. J\o{}rgensen (\url{jakj@dtu.dk}). This work was supported by The Villum Foundation (grant no.\ 25893).
J.S.J. would like to thank the Isaac Newton Institute for Mathematical Sciences for support and hospitality during the programme ``Rich and Nonlinear Tomography -- a multidisciplinary approach'' when work on this paper was undertaken. 
This work was supported by EPSRC Grant Numbers EP/P02226X/1, EP/T026677/1, EP/T026693/1 and EP/R014604/1. 
This work was partially supported by a grant from the Simons Foundation (J.S.J.). This work made use of computational support by CoSeC, the Computational Science Centre for Research Communities, through CCPi and CCP-SyneRBI. The authors are grateful to Emil Sidky for valuable discussions on directional total variation.}
\end{abstract}

\section{Introduction} \label{sec:introduction}
\begin{figure}[tb]
    \centering
    \includegraphics[width=0.95\textwidth]{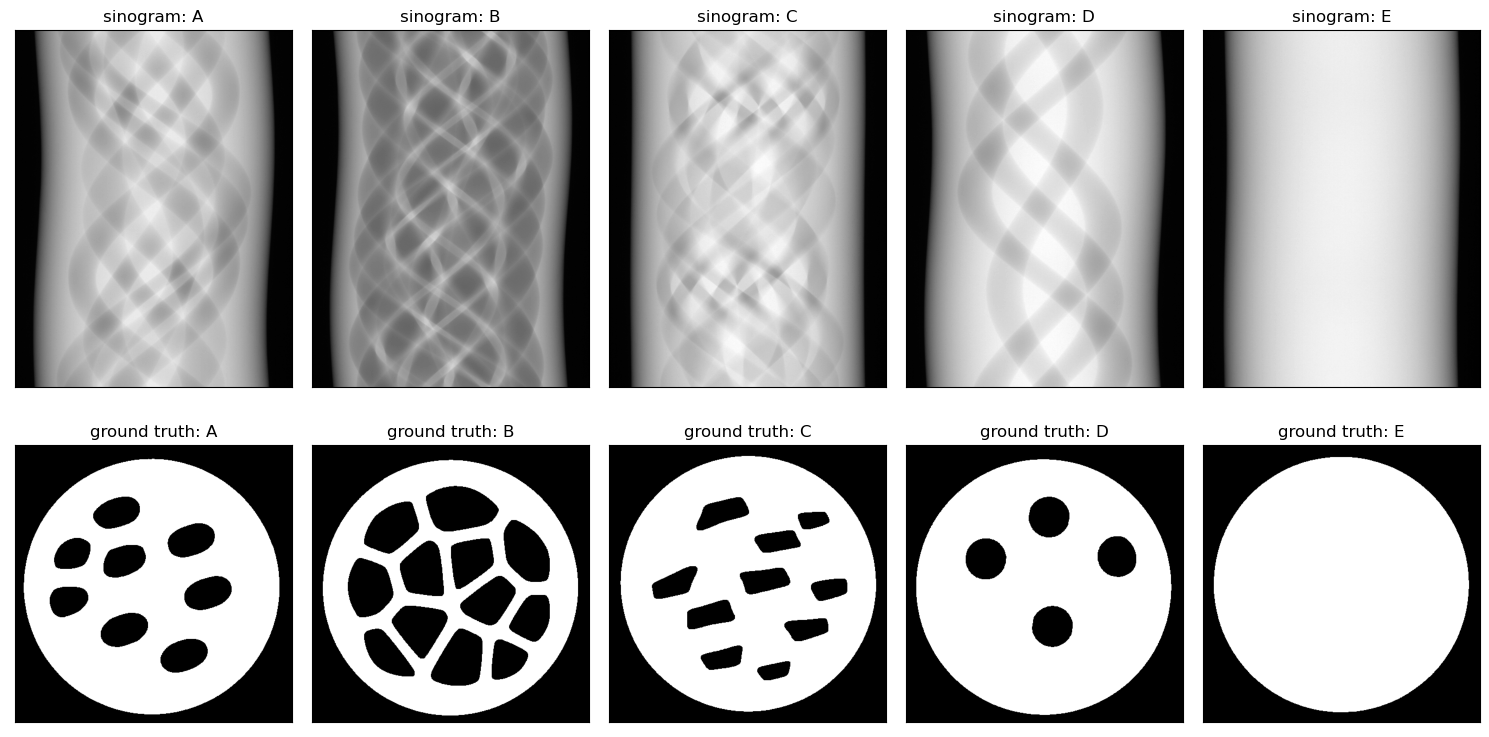}
    \caption{The 5 training datasets provided. The top row shows the full 360\textdegree\ sinograms A-E. The bottom row shows the ground truth reconstructions A-E.}
    \label{fig:test_data_initial} 
\end{figure}

The Finnish Inverse Problem Society’s 2022 Helsinki Tomography Challenge (HTC)\footnote{\url{https://www.fips.fi/HTC2022.php}} tasked participants with developing algorithms for reconstructing the shapes of 2D phantoms from limited-angle tomographic data. The challenge provided 5 training datasets and their ground-truth reconstructions, shown in figure \ref{fig:test_data_initial}. These datasets were used to develop the submitted algorithms by sampling different ranges of angles from each. In the evaluation phase, the submitted algorithms were tested on unseen limited-angle datasets with angular ranges between $90^\circ$ and $30^\circ$.

The developers of the Core Imaging Library (CIL) formed a team and developed and entered five algorithms, with the performance of one algorithm acquiring the third place in the competition. 
This article presents these five algorithms, with a particular focus on the best-performing one\footnote{\url{https://github.com/TomographicImaging/CIL-HTC2022-Algo2}}.

The algorithms were developed using CIL \cite{Jorgensen2021, Papoutsellis2021}, an open-source python package for pre-processing and reconstructing challenging tomographic data, developed by CCPi, the Collaborative Computational Project in tomographic imaging\footnote{\url{https://ccpi.ac.uk/cil}}. CIL provides a number of different reconstruction algorithms, and has a “plug and play” structure which allows to easily create new algorithms using a variety of building blocks, including data fidelity terms, constraints and regularizers.
All of the algorithms entered by the CIL team use an iterative method to solve an optimization problem, balancing a data fidelity term, one or more regularization term using variants of Total Variation (TV) as well as a tailored set of lower and upper bound constraints.

The algorithms submitted by the CIL team were the only ones taking a model-based reconstruction approach. The remaining teams followed a deep learning approach, by generating a large amount of synthetic phantom training data, which was feasible due to the small size of the data. In particular, the first-placed team\footnote{\url{https://github.com/99991/HTC2022-TUD-HHU-version-1}} trained a Convolution Neural Network (CNN) to predict a reconstructed image from limited-angle data and the second-placed team\footnote{\url{https://github.com/alexdenker/htc2022_LPD/tree/master}} used a primal-dual unrolling method based on \cite{ChambollePock}. Although the power and effectiveness of deep learning methods are undeniable, we demonstrate that with the right pre-processing tools and modelling we can deliver exceptional model-based reconstruction results.

This paper is organized as follows. \Cref{sec:literature} gives an overview of model-based regularization approaches for limited-angle tomography, \cref{sec:preprocessing_segmentation} describes the data pre-processing (prior to reconstruction) steps of the presented method, as well as the image segmentation method. \Cref{sec:reconstruction} describes and illustrates the components of the presented reconstruction method, \cref{sec:results} provides experimental results and comparisons, before a discussion incl. future work is given in \cref{sec:discussion} and the work concluded in \cref{sec:conclusion}.

We use the following notation. Images and sinograms are written as vectors $\im$ and $\sino$ in bold lower-case, linear operators like the CT system matrix as bold upper-case $\sysmat$, a vector of all zeros or all ones as $\zerovec$ and $\onevec$, with the size given by context. We use $\maskvec$ to denote a binary vector representing a mask. Convex functionals used in optimization as $F$, $G$, etc., regularization parameters as greek $\omega$, $\alpha$, $\beta$. Spatial horizontal and vertical dimensions are $x$ and $y$ respectively. Linear attenuation coefficent is denoted $\mu$ and the specific value used for the acrylic disks in this work is denoted $\mu_\text{A}$.

\section{Model-based limited-angle CT reconstruction}
\label{sec:literature}

Computed Tomography (CT) is a fundamental imaging technique to reconstruct a two- or three-dimensional object from acquired projections \cite{Hansen2021}. In a conventional scanning, measured projections
cover a full $180^\circ$ view of the sample for a  parallel-beam geometry and $360^\circ$ view for cone-beam geometry. However, there are cases where full view scanning is not possible, for example in dental CT, in breast tomosynthesis or in-situ imaging where experimental rigs may partially block the view. 

In the context of CT reconstruction, two primary methods are commonly used in practice: analytic inversion and iterative methods. For analytic reconstruction, we have Filtered Back Projection (FBP) for parallel-beam geometries and the Feldkamp, Davis and Kress (FDK) method for cone-beam geometries. Usually, these are quite efficient and accurate but they rely on the assumption that the acquired data is well sampled, including a full angular range. Reconstruction from a limited angular range of projections is a very challenging and highly ill-posed problem that for FBP and FDK results in dramatic limited-angle artifacts including streaks and directional blurring \cite{Frikel2013}.

Iterative reconstruction methods employ a fully discretized linear imaging model 
\begin{align}
    \sysmat \im = \sino, \label{eq:linearproblem}
\end{align}
where $\im \in \mathbb{R}^n$ is the unknown image to be determined having $[\im]_j$ for $j=1,\dots, n$ as pixel values, $\sysmat \in \mathbb{R}^{m\times n}$ is the forward operator or system matrix and $\sino \in \mathbb{R}^m$ the observed log-transformed attenuation data. Each row of $\sysmat$ corresponds to a discretized version of the Lambert-Beer law for attenuation of the $i$th X-ray passing through the object along the line $L_i$, 
\begin{align}
    I_i = I_0 \exp( - a_i )\qquad \text{for}\qquad a_i = \int_{L_i} \mu(x,y) d\ell,\label{eq:beerlambert}
\end{align}
where $\mu(x,y)$ is the linear attenuation coefficient at position $(x,y)$, $I_0$ is the incident X-ray intensity and $I_i$ the measured intensity after passing through the object, and $a_i$ is called the absorption of ray $i$.

We focus here on the class of variational or regularized iterative reconstruction methods, which have been shown to produce better images with reduced noise and suppressed limited angle artifacts, by taking into consideration the noise characteristics of the data.\cite{Beister2012}. In this framework we formulate and numerically solve an optimization problem to combine the physical model in a \emph{data fidelity term} $D(\im)$, any prior information available using one or more \emph{regularization terms} $R(\im)$ as well as any constraints $\im \in \mathcal{C}$, where $\mathcal{C}$ is normally a convex set. We write
\begin{equation}
\underbrace{\hat{\bm{u}}}_\text{solution} = \quad \argmin_{\bm{u}} \quad\,\,\underbrace{D(\bm{u})}_\text{data fidelity} \,\,+\,\,\underbrace{R(\bm{u})}_\text{regularizer}    \qquad \text{subject to} \qquad \underbrace{\im \in \mathcal{C}}_\text{constraint}.
\label{eq:general_objective}
\end{equation}
The first term in \eqref{eq:general_objective} measures the discrepancy between the noisy tomography data and the reconstructed image, the second term enforces some prior assumption such as the signal being smooth, piecewise constant or similar, while the constraint can for example be used to ensure all pixel values are nonnegative. One or more of the terms include a multiplicative regularization parameter to govern the influence of each of the terms to balance between fitting the data and enforcing regularity on the image.

In the last three decades, a variety of iterative reconstruction methods have been proposed in the context of limited-angle tomography. They focus mainly on a) the actual implementation of the algorithm, b) the objective function to be minimized and c) tuning strategies of the regularization parameter. We give a brief overview focused on choices of data fidelity terms and regularizers we employed in our five algorithm challenge submissions.

\subsection{Data fitting term}
\label{subsec:data_fitting_term}
In general, the choice of data fidelity can be motivated from the distribution of noise on the data in the model \cref{eq:linearproblem}. CT data is often modelled as a Poisson process, i.e., data is subject to Poisson noise, however for relatively high photon counts it is reasonable to use a Gaussian approximation \cite{Bouman1993}. From a maximum-likelihood perspective this becomes a least-squares problem and in the simplest case one simply chooses the following, which is readily available in CIL:
\begin{align}
    D(\im) = \| \sysmat \im - \sino\|_2^2,
\end{align}

It is worth mentioning that for low-count applications of different tomography modalities, signal dependent noise modelling is more appropriate. In this case, we could use other data fitting terms that are already available through CIL. For example in low-dose CT \cite{Tian2011} and medical imaging applications,  such as Positron Emission Tomography (PET) and Single Photon Emission Computed Tomogrpahy (SPECT),  one can use the Kullback-Leibler (KL) divergence  \cite{Resmerita2007}. We refer the  reader to \cite{Brown2021}, where we employ this fidelity term and combine CIL, as an optimization engine, together with the \emph{Synergistic Image Reconstruction Framework (SIRF)}, \cite{Ovtchinnikov2020}, that implements an accurate forward model for PET reconstruction. In addition, we offer an approximation of the KL divergence by a second order Taylor expansion, namely the Penalized Weighted Least Squares (PWLS), \cite{Sauer1993}.

\subsection{Regularization terms}
\label{sec:TV}
The total variation (TV) semi-norm which is equivalent to the $\ell^{1}$-norm of the gradient has been used extensively for limited-angle tomography reconstruction. Assuming that the desired reconstruction has piecewise constant structures, the authors in \cite{Sidky2006} minimize the isotropic TV of $\bm{u}\in\mathbb{R}^n$, defined in 2D as 
\begin{equation}
    \mathrm{TV_{i}}(\bm{u}) = \|\bm{D} \bm{u}\|_{2,1} = \sum_{j=1}^n\bigg[\sqrt{(\bm{D}_{x} \bm{u})^{2} + (\bm{D}_{y} \bm{u})^{2}}\bigg]_j
    \label{isotropicTV}
\end{equation}
where $\bm{D}=(\bm{D}_{x}, \bm{D}_{y})$ is the discrete gradient operator, approximated with forward differences under Neumann boundary conditions.
With this regularizer, noise and streak artifacts are removed for limited-angle applications. Later, this method was improved in \cite{Sidky2008, Bian2010}, using the Algebraic Reconstruction Technique (ART), based on \cite{Andersen1984}. Due to the loss of contrast and staircasing artifacts that TV regularization produces, a better reconstruction can be obtained using local information of the image \cite{Tian2011, Liu2012} or high-order extensions of TV such as Total Generalized Variation (TGV), \cite{Niu2014}. Another important work which incorporates prior image information, such as FBP reconstruction, is the Prior Image Constrained Compressed Sensing (PICCS) method, \cite{Chen2008}. Such methods assume that the difference on the gradient between the prior and the reconstructed images is small, achieving significant improvements for cardiac CT reconstructions. A similar approach in which edge information from a reference image is propagated into the reconstruction process is presented in \cite{Papoutsellis2021}. 

Another successful approach  is   to decompose the magnitude of the gradient and separately penalize the variations along the image dimensions using the anisotropic TV 
\cite{Esedolu2004} defined as 
\begin{equation}
    \mathrm{TV_{a}}(\bm{u}) = \|\bm{D} \bm{u}\|_{1,1} = \sum_{j=1}^n\bigg[|(\bm{D}_{x}\bm{u})| + |(\bm{D}_{y}\bm{u})|\bigg]_j = \mathrm{TV}_{x}(\bm{u}) + \mathrm{TV}_{y}(\bm{u})
    \label{anisotropicTV}
\end{equation}

In limited-angle tomography, where some projections  are missing we expect to have some streak artifacts in some directions and to preserve edges and image structures in other directions. Using \eqref{isotropicTV}, the reconstruction contains contributions from directions with both sharp edges and image boundaries that are blurred. A separate penalization along each direction can act as an additional prior knowledge coming from the angular range of the scanning, see for instance \cite{Chen2013, Wang2017}. 

Recently, the authors in \cite{Zhang2021} performed a numerical study on breast and bar phantoms comparing the isotropic and anisotropic TV regularization for limited-angle CT reconstruction. The proposed method called Direction Total Variation (DTV) is solved differently than \cite{Sidky2008}. Here, the fidelity term is minimized and the constraints are imposed on the two image gradients under different weights. Our highest-scoring algorithm \ref{Iso1DAnisoTV} described below follows a similar setup. Lastly, we would like to mention two reconstruction methods that combine isotropic and anisotropic regularization at the same optimization problem. One is the weighted difference between isotropic and anisotropic TV, i.e., $\|\bm{D}\bm{u}\|_{1,1} - \|\bm{D}\bm{u}\|_{2,1}$ presented in \cite{Lou2015} and the other is the ratio, i.e., $\|\bm{D}\bm{u}\|_{1,1} / \|\bm{D}\bm{u}\|_{2,1} $ recently proposed in \cite{Wang2020}. We focused on the weighted difference case in CIL.

\section{Pre-processing and segmentation methods}
\label{sec:preprocessing_segmentation}

In this section we describe all steps applied before the reconstruction, starting with the loading of data into CIL. We then describe the sequence of pre-processing steps that we applied and finally the image segmentation method we used to segment the image into phantom and background pixels.

\subsection{Loading the data}
We start by defining the geometry of the CT system as a CIL object. In the code snippet below we create \texttt{data}, a CIL acquisition data object which holds the measured sinogram values as well as the description of the system acquisition geometry, \texttt{ag}. The acquisition geometry is a fan-beam CT, with 560 pixels of $0.2\mm$. The source to object distance $d_\mathrm{SO} = 410.66\mm$ and the object to detector distance $d_\mathrm{OD} = 143.08\mm$.
CIL's image geometry, \texttt{ig}, is the region of voxels that we will compute the CT volume for. The challenge specifies this as $512 \times 512$ voxels at full resolution, i.e., with image pixel size equal to the the detector pixel size divided by the geometric magnification $(d_\mathrm{SO} + d_\mathrm{OD})/d_\mathrm{SO}$.

\texttt{show\_geometry()} is a utility function that shows the system geometry. The output of this can be seen in \cref{fig:padding}. \texttt{show2D()} is a utility function which displays a slice of data in the CIL data container, either a reconstructed image or as here a sinogram:

\begin{center}
\begin{tcolorbox}[
    enhanced,
    attach boxed title to top center={yshift=-2mm},
    colback=darkspringgreen!20,
    colframe=darkspringgreen,
    colbacktitle=darkspringgreen,
    title=Creating the CIL geometry and data,
    text width = 12.7cm,
    fonttitle=\bfseries\color{white},
    boxed title style={size=small,colframe=darkspringgreen,sharp corners},
    sharp corners,
]
\begin{minted}{python}

ag = AcquisitionGeometry.create_Cone2D( source_position=[0,-410.66], 
                                        detector_position=[0,143.08])
ag.set_panel(num_pixels=560, pixel_size=0.2)
ag.set_angles(-angles_input, angle_unit='degree')

data = AcquisitionData(sinogram_data, geometry=ag)

ig = ag.get_ImageGeometry()
ig.voxel_num_x = 512
ig.voxel_num_y = 512

show_geometry(ag, ig)
show2D(data)
\end{minted}
\end{tcolorbox}
\end{center}

\subsection{Renormalization}

\begin{figure}[tb]
    \centering
    \includegraphics[width=0.9\textwidth]{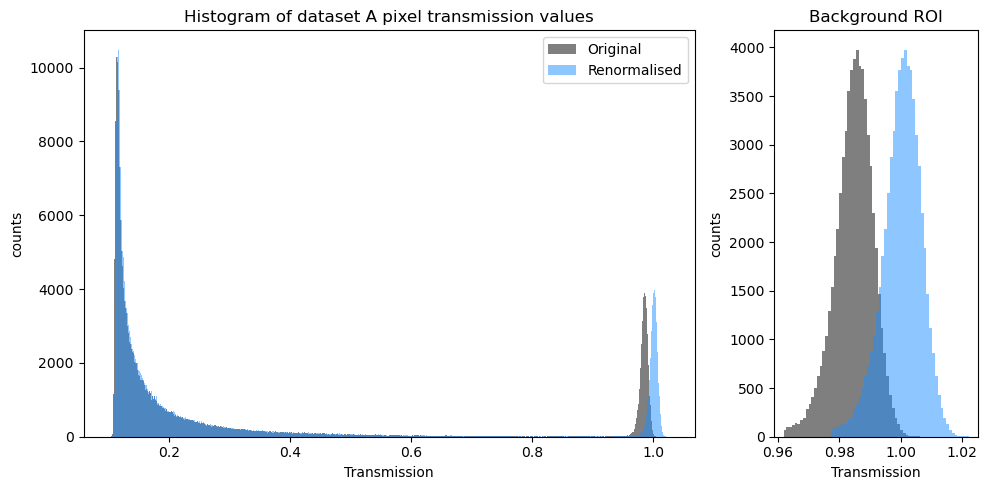}
    \caption{Left: Histogram of the pixel transmission values before and after re-normalization. Right: Zoom around 1.0.}
    \label{fig:renormalisation} 
\end{figure}

\Cref{fig:renormalisation} shows a histogram of the transmission values (i.e. $I_i/I_0$ for $i=1,\dots,m$) of dataset A. We expect to see a peak at $1$ corresponding to rays that do not intersect the sample and so are not attenuated. The mean of the background peak is below $1$, at approximately $0.98$. A possible cause is that source characteristics changed between the CT and the flat-field data collection steps. This shifted background can present a difficulty for beam-hardening correction (\cref{sec:BHC}) and optimization problems as it is interpreted as attenuating material that has to be distributed among the reconstructed voxels.

In order to renormalize the data, since the challenge data is provided as absorption data, we convert this to transmission data using \cref{eq:beerlambert}.
 We then divide through by the value of the background peak which re-scales the pixel intensities to the new range, and then convert the data back to absorption. The original and shifted transmission data can be seen in \cref{fig:renormalisation}. In CIL, these steps can be performed using the following code:

\begin{center}
\begin{tcolorbox}[
    enhanced,
    attach boxed title to top center={yshift=-2mm},
    colback=darkspringgreen!20,
    colframe=darkspringgreen,
    colbacktitle=darkspringgreen,
    title=Renormalization,
    text width = 12.7cm,
    fonttitle=\bfseries\color{white},
    boxed title style={size=small,colframe=darkspringgreen,sharp corners},
    sharp corners,
]
\begin{minted}{python}

data_trans = AbsorptionTransmissionConverter()(data_abs)
data_trans_renorm = data_trans / 0.98438
data_abs_renorm = TransmissionAbsorptionConverter()(data_trans_renorm)
\end{minted}
\end{tcolorbox}
\end{center}

\subsection{Zero padding}
    
\begin{figure}

    \setlength\tabcolsep{0pt}
    \begin{tabular*}{\linewidth}{@{\extracolsep{\fill}} ccc }
    
        \includegraphics[height=0.245\linewidth]{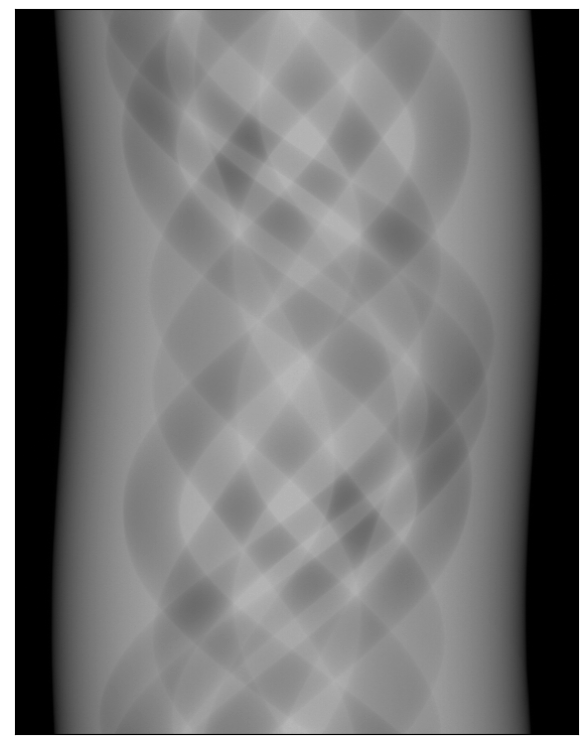} &
        \includegraphics[height=0.245\linewidth]{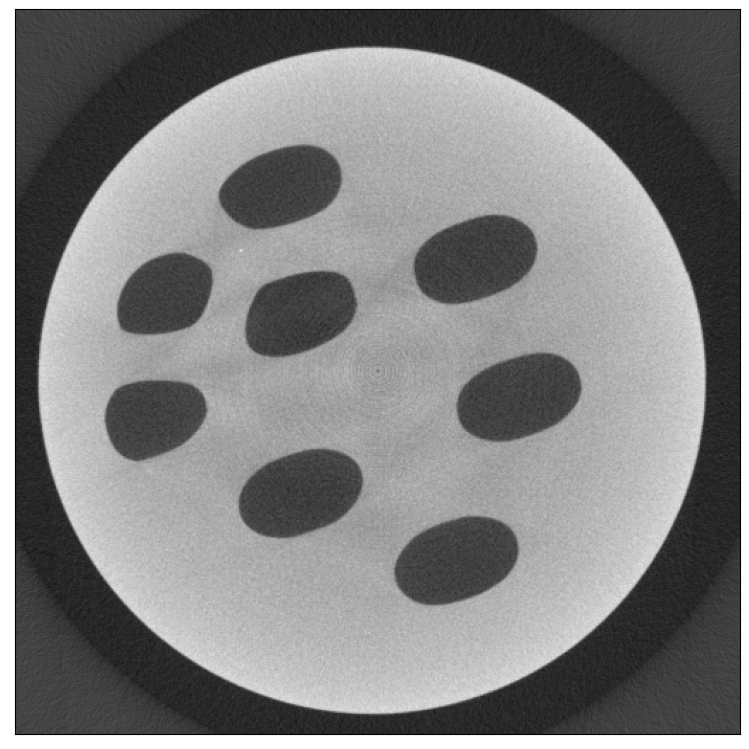} &
        \includegraphics[height=0.245\linewidth]{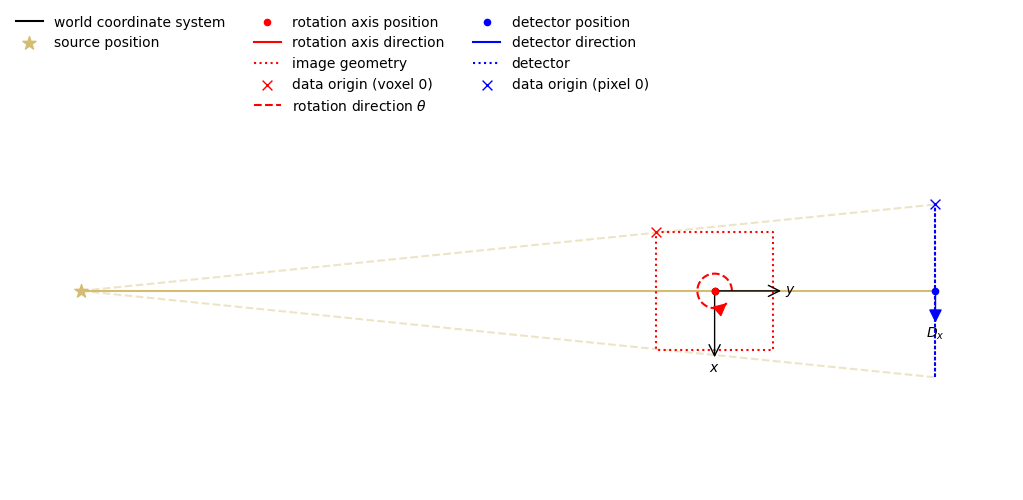}  \\
        \includegraphics[height=0.245\linewidth]{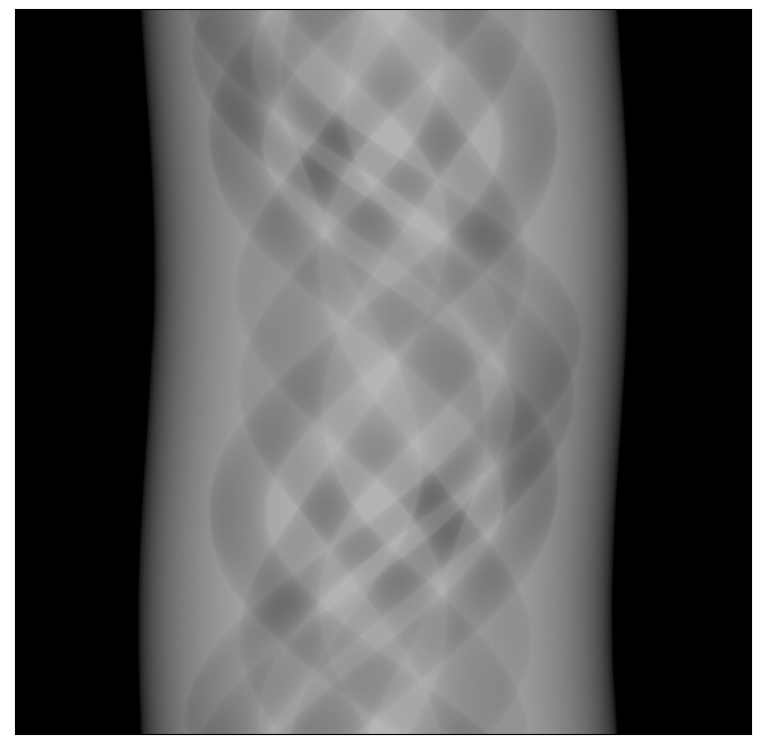} &
        \includegraphics[height=0.245\linewidth]{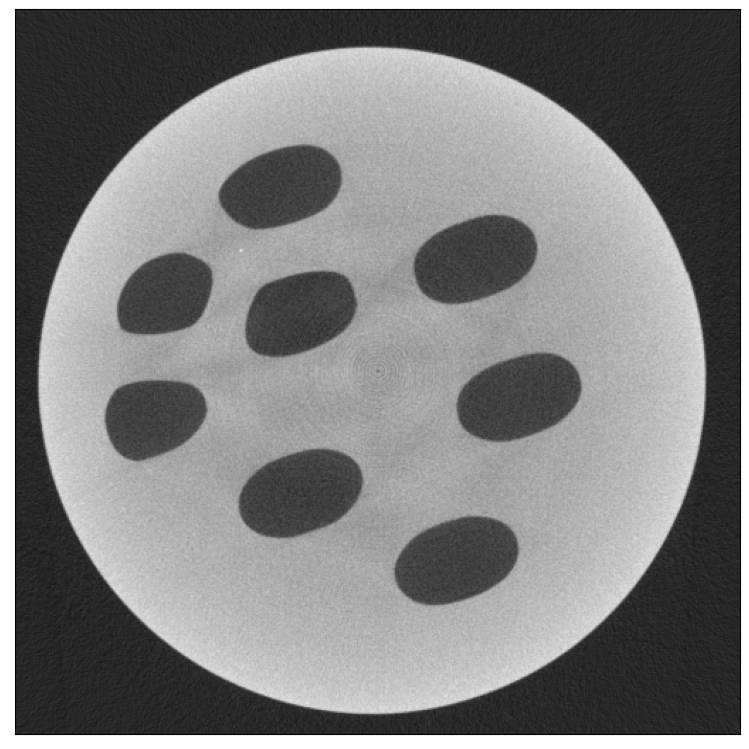} &        \includegraphics[height=0.245\linewidth]{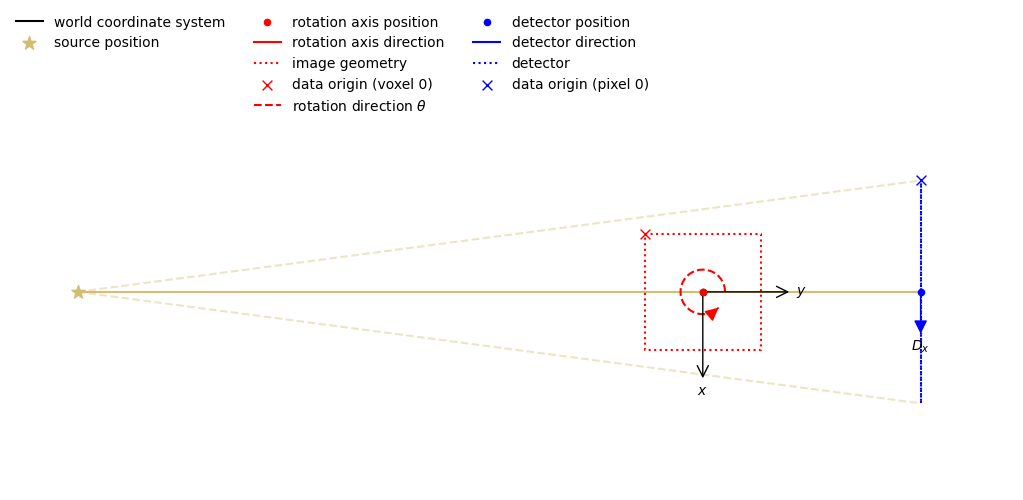}  \\

    \end{tabular*}

    \caption{The sinogram, image data of the FDK reconstruction and the system geometry for the original (top) and zero-padded (bottom) data.}
    
    \label{fig:padding} 
\end{figure}

 If doing an FDK reconstruction (\cref{sec:fdkrecon}) the resulting image shows a brighter ring of voxels around the outside of the sample; see the top row of \cref{fig:padding} for data, reconstruction and corresponding geometry. The ring is due to the acquisition geometry extending outside the x-ray fan-beam at certain angles. If reconstructing with a single back-projection it is common practice to mask this ring after reconstruction, however, for iterative methods it is helpful to address this in the input data.

By simply padding the data with zeros, but keeping the original image geometry we can suppress this ring in the back-projection and force the full background region to zero. Here we pad by 86 pixels on each side of the data. This now means the image geometry is inside the now wider fan-beam at every angle, and the background region of the reconstruction is now uniform. The bottom row of \cref{fig:padding} shows the zero padded data and the corresponding FDK reconstruction and system geometry.

The CIL \texttt{Padder} processor can pad using all common modes (i.e. constant value, reflecting, edge extending). The geometry will be modified along with the data and so stay consistent even when the border is asymmetric. Here we show the code used to pad the data and show the padded sinogram:

\begin{center}
\begin{tcolorbox}[
    enhanced,
    attach boxed title to top center={yshift=-2mm},
    colback=darkspringgreen!20,
    colframe=darkspringgreen,
    colbacktitle=darkspringgreen,
    title=Zero padding,
    text width = 12.7cm,
    fonttitle=\bfseries\color{white},
    boxed title style={size=small,colframe=darkspringgreen,sharp corners},
    sharp corners,
]
\begin{minted}{python}

data_padded = Padder.constant(pad_width=86, constant_values=0)(data)
show2D(data_padded)
\end{minted}
\end{tcolorbox}
\end{center}

\subsection{Beam-hardening correction}
\label{sec:BHC}

\begin{figure}[tb]
    \begin{tabular}{ cc }
        \includegraphics[height=0.25\textwidth]{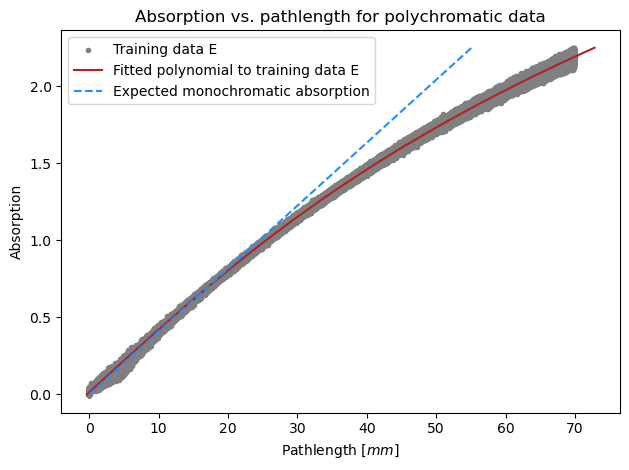}&
        \includegraphics[height=0.25\textwidth]{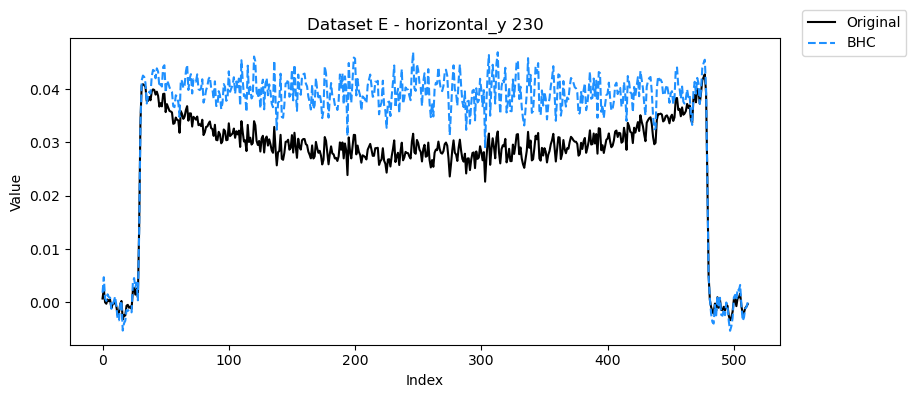}\\
        \includegraphics[height=0.25\textwidth]{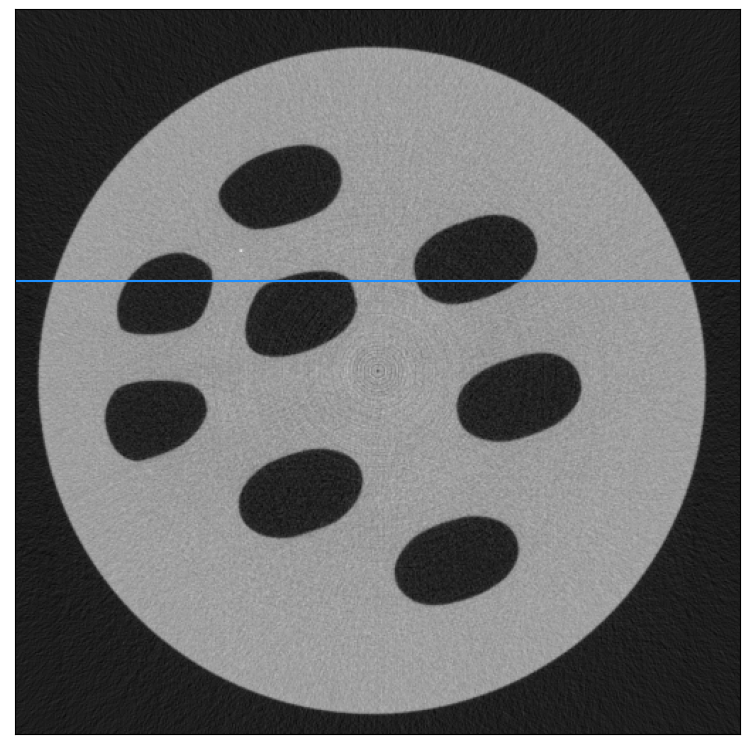}&
        \includegraphics[height=0.25\textwidth]{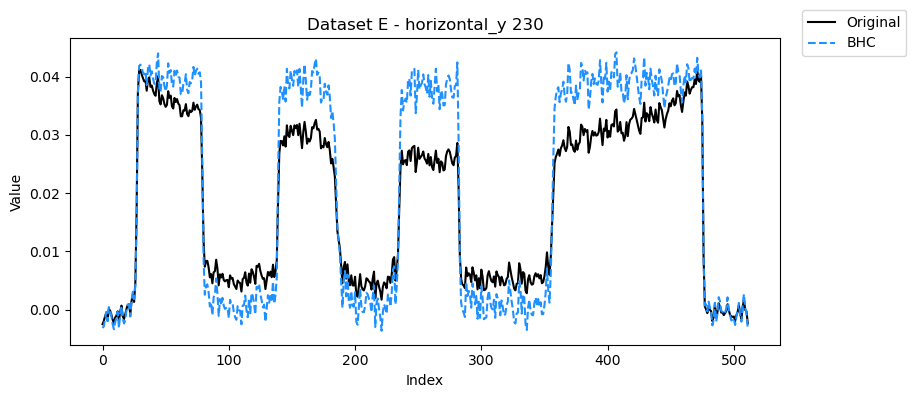}\\
    \end{tabular}

    \caption{Top Left: Absorption vs. path length for training dataset E. This shows the polynomial fit to pre-processed data and the estimated monochromatic correction. Top Right: Line profiles through the FDK reconstructions of dataset E with and without beam-hardening correction. Bottom left: FDK reconstruction of dataset A after beam-hardening correction has been applied. Bottom right: Line profiles through the FDK reconstruction of dataset A with and without beam-hardening correction.}

    \label{fig:BHC} 
\end{figure}

The attenuation of a material has an energy dependence which results in lower energy x-rays being preferentially absorbed and in the case of polychromatic beams, the mean energy of a beam to increase. This is known as beam-hardening.

The physical model \cref{eq:linearproblem} for reconstruction neglects energy dependence of the measured data and a typical reconstruction treats the source as if it was monochromatic. This leads to artefacts in the CT volume that present as cupping\cite{Brooks_1976} and streaking artefacts\cite{Joseph_1978}. Cupping artefacts will make a uniform sample appear denser towards its edges, streaking artefacts appear to bridge between areas of material. These can be seen visually in the reconstructions shown in  \cref{fig:padding} as well as in the line profiles in \cref{fig:BHC}.

For a single-material sample it is possible to estimate what the absorption of a monochromatic beam would have been from the polychromatic absorption data\cite{Herman_1979}. This correction can be applied to the data creating a pseudo-monochromatic dataset which is then artifact-free when reconstructed. This linearization of the data is well established and has been shown to work well for single material datasets\cite{Herman_1983}. The method makes use of the fact that for a single-material sample, the Beer-Lambert law \cref{eq:beerlambert} simplifies such that absorption along line $i$ can be written as simply as the product $a_i = \mu p_i$, where $\mu$ is the constant linear attenuation of the material and $p_i$ is the path length through the sample.

We start by using training dataset E, as this is a solid disk containing the longest path-lengths of any of the samples. We first perform an FDK reconstruction of the sample, we segment this and forward-project it in order to get the path lengths $p_i$ through the material of each ray $i=1,\dots,m$. The top-left image in \cref{fig:BHC} shows the absorption vs. path lengths. We perform a 3rd degree polynomial fit to this data. This fit is shown in red and can be seen to approximate the absorption vs. path lengths well. We want to map this data on to a linear model. The linear model is the monochromatic estimation, shown in blue. This monochromatic estimate passes through the origin, with the gradient tangential to the polynomial fit at small path-lengths. Dividing the polynomial coefficients by this selected gradient we obtain a new polynomial that encodes this mapping. This is then applied to the polychromatic absorption data to create a pseudo-monochromatic absorption dataset. A line profile through the reconstruction of the pseudo-monochromatic dataset is show in the top-right of \cref{fig:BHC}. We can see the cupping artifacts have been corrected.

The gradient of the monochromatic estimation relates to the linear attenuation coefficient of the sample. The gradient we have chosen corresponds to $\mu = 0.0409 \mm^{-1}$. We can use the NIST X-ray attenuation data \cite{NIST} to see that this corresponds to a monochromatic energy of approximately $24.7 \mathrm{keV}$ for acrylic (PolyMethyl MethAcrylate). This energy is close to the expected micro CT source spectrum mean-energy.

In principle, one may apply the same beam-hardening correction procedure to all new datasets. However, this method requires an estimation of path lengths through the sample, which is not feasible when only limited-angle data is available. This is because the path lengths are determined using a forward projection of a reconstructed dataset, which will be poor for limited-angle data. Instead we apply the coefficients found from dataset E directly to each dataset in the challenge.  This makes assumptions that the source spectrum remains constant throughout the data acquisitions and that the path lengths fall within our fitted range.

Applying the fitted coefficients to training data A and performing an FDK reconstruction we see in \cref{fig:BHC} a noticeable reduction in artifacts once the beam-hardening correction has been applied, compare bottom row of \cref{fig:padding}. The material density distribution across the disk now stays constant, and the contrast between the acrylic and the holes in the disk has been recovered.

\subsection{Segmentation and assessment}
\label{sec:segmentation}

\begin{figure}[tb]
    \centering
    \includegraphics[width=0.9\textwidth]{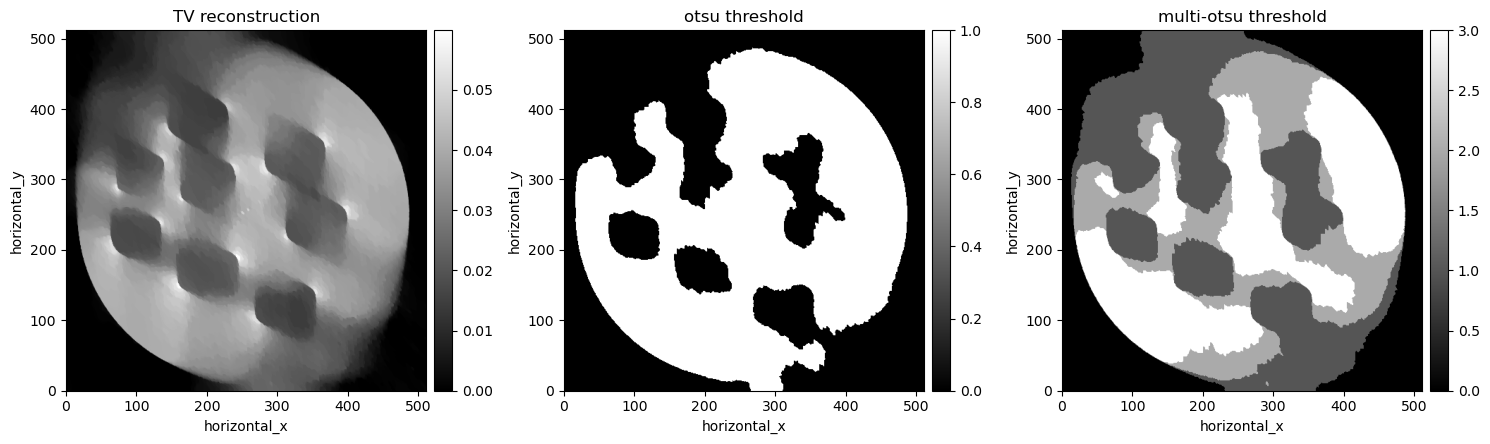}
    \includegraphics[width=0.9\textwidth]{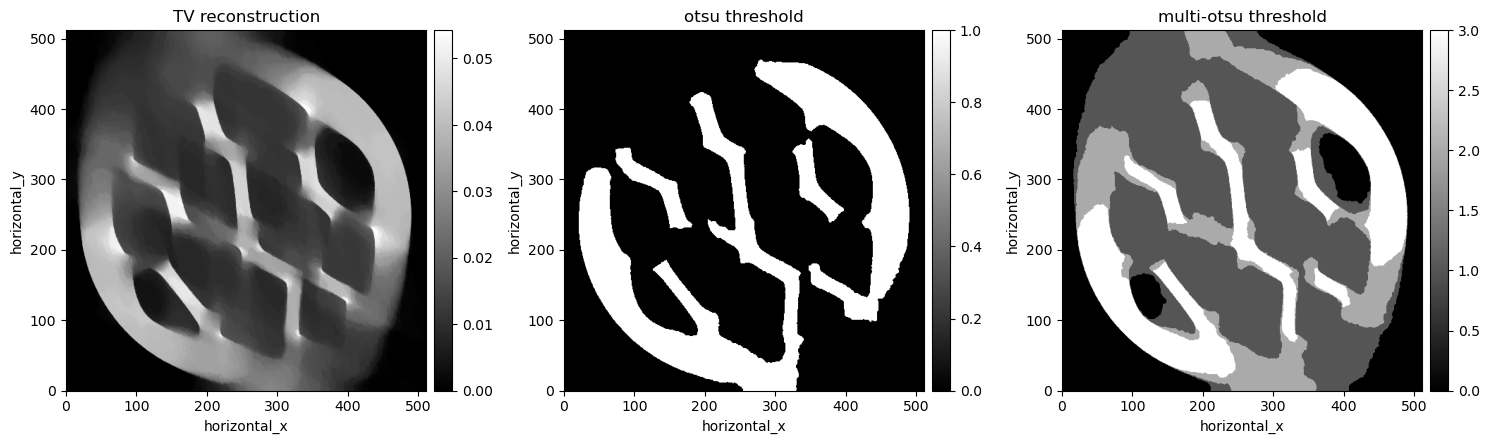}
    \caption{Performance comparison of Otsu and multi-Otsu segementation on $60^\circ$ data. Left: TV reconstruction. Center: Otsu segmented reconstruction. Right: multi-Otsu segmented reconstruction. Top: Training data A. Bottom: Training data B.}
    \label{fig:seg-otsu} 
\end{figure}

The challenge rules state the following scoring procedure: 
``\emph{The reconstructions will be assessed quantitatively, comparing the reconstructed binary image with the ground truth binary image, assigning a numeric score. The score is based on the confusion matrix of the classification of the pixels between empty (0) or material (1). [\ldots] A score of $+1$ (best) represents a perfect reconstruction, $0$ no better than random reconstruction, and $-1$ (worst) indicates total disagreement between reconstruction and ground truth.}'' Code to compute this score was provided.

None of our reconstruction algorithms (\cref{sec:reconstruction}) directly output a binary solution required for the scoring procedure, therefore segmentation is evidently an important step. The segmentation defines which pixels are classified as material and which are background, which in turn determines the score of each algorithm. Reconstructing the full angular range of a single material sample gives us a result that is trivial to segment. As the difficulty of the challenge increases we have more uncertainty on material labels and the segmentation can have a high impact on the final score. 

We looked at 60\textdegree\ of data reconstructed with only the TV-regularized \cref{{eqn:TV}}. On these reconstructions we compared Otsu segmentation with multi-Otsu segmentation \cite{Otsu} as implemented by Python package scikit-image\cite{van2014scikit}. \Cref{fig:seg-otsu} compares a single-threshold segmentation with a 3-threshold segmentation on training datasets A and E. In the multi-Otsu case we then take the center threshold setting the two highest attenuating segments as acrylic, and the two lowest as air.

On dataset B (lower \cref{fig:seg-otsu}) the multi-threshold produces a visually better segmentation of the data than is achieved with the single threshold. The challenge score based on these reconstructions is $0.732$ in the multi-Otsu case and $0.707$ in the Otsu case. In this instance multi-Otsu raises the score of the result, despite using the same reconstruction methods. Based on preliminary work we selected the 3-threshold segmentation in all of our algorithms. However, looking at the same reconstruction for dataset A (upper \cref{fig:seg-otsu}), multi-Otsu and Otsu  achieve visually very similar results and the challenge score based on these reconstructions is $0.778$ in the multi-Otsu case and $0.792$ in the Otsu case. This shows that there are instances where a simple threshold may have been a better choice.

The segmentation is task dependent and more sophisticated methods are an area of potential great improvement but fall beyond the scope of this work.

\section{Reconstruction}
\label{sec:reconstruction}

In this section we describe the relevant prior information we identified about the challenge data and how we incorporated it into optimization problems to compensate for the limited-angle data. 

We illustrate each step by means of data from the training set, mainly sample A, presenting a series of images comprising the obtained reconstruction with 50 degrees data, the segmentation and the difference between the obtained segmentation and the ground truth to highlight misclassified pixels, figures \ref{fig:isotropic_tv},  \ref{fig:isotropic_tv_disc} and \ref{fig:anisotropic_tv}, alongside with the score used by the organizers in the assessment of the algorithms. The ground truth is the segmentation provided by the organizers, which is obtained from the full data FDK reconstruction.

\subsection{Baseline: FDK reconstruction} \label{sec:fdkrecon}
\begin{figure}[tb]
    \centering\includegraphics[width=0.9\textwidth]{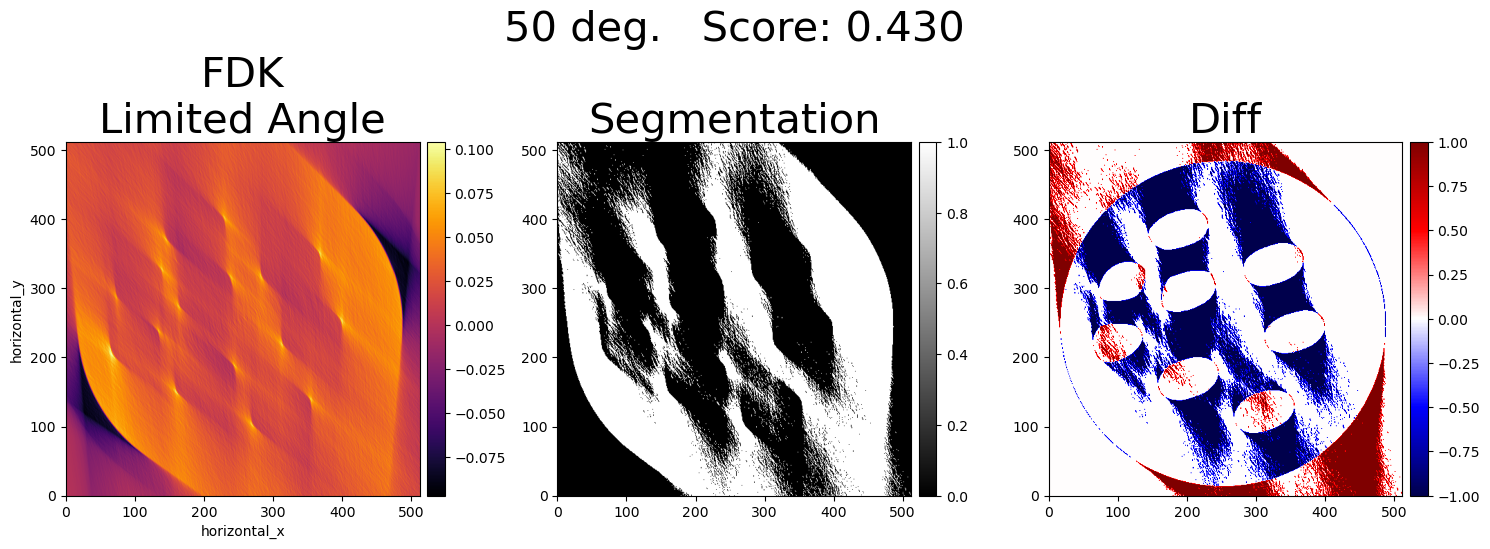}
    \caption{Reconstruction with FDK of the training sample A with 50$^\circ$ data. Left: reconstruction. Center: segmentation. Right: Difference between segmentation and ground truth segmentation provided by the organizers. Correctly classified pixels are white, true background pixels misclassified as acrylic are red, and true acrylic pixels misclasssified as background are blue.}
\label{fig:fdk_score}
\end{figure}

With the data pre-processing and segmentation methods set up, the first port of call for reconstruction is filtered back-projection (FBP) using the standard Feldkamp-Davis-Kress (FDK) method \cite{Feldkamp:84} here just applied to the 2D fan-beam geometry. In CIL this can done simply as:
\begin{center}
\begin{tcolorbox}[
    enhanced,
    attach boxed title to top center={yshift=-2mm},
    colback=darkspringgreen!20,
    colframe=darkspringgreen,
    colbacktitle=darkspringgreen,
    title=FDK reconstruction,
    text width = 12.7cm,
    fonttitle=\bfseries\color{white},
    boxed title style={size=small,colframe=darkspringgreen,sharp corners},
    sharp corners,
]
\begin{minted}{python}

reconstruction = FDK(data, ig).run()
show2D(reconstruction)
\end{minted}
\end{tcolorbox}
\end{center}
using GPU-acceleration either by ASTRA \cite{vanAarle2016} or TIGRE \cite{Biguri2016} backends. The resulting FDK reconstruction for the $50^\circ$ case of the training sample A, its segmentation as well as the difference image with the ground truth segmentation provided by challenged organizers are seen in \cref{fig:fdk_score}. As expected, there are prominent limited-angle artifacts with strong directional blurring from upper left to bottom right. The segmentation is poor which is highlighted by the difference image in which white shows correctly classified pixels, red background pixels misclassified as acrylic and blue acrylic pixels misclassified as background. The score is 0.430.

\subsection{The prior information available}
The basis of model-based iterative reconstruction consists of a data fitting term, which encodes the physical model, and of the prior information about the image we are seeking to reconstruct, see \ref{subsec:data_fitting_term}.
We know from the challenge description that the targets are homogenous acrylic disk phantoms of 70 mm in diameter, with holes of varying
shapes made with a laser cutter.
Therefore, the prior knowledge (PK) that we had on our samples and we encoded in the creation of our algorithm is the following:
\begin{enumerate}[label=PK\arabic*]
    \item \label{list:pk-homogeneous} The samples are effectively binary-valued, consisting of a single homogeneous material with sharp edges wrt. the background air.
    \item \label{list:pk-disc} Their outer shape is approximately that of a disk, and we expect zero attenuation outside the object and a constant value of $\mua = 0.0409$ mm$^{-1}$ inside the object.
    \item \label{list:pk-directional} As can be seen from \cref{fig:fdk_score} some edges are reconstructed correctly, while some are severely smeared. According to microlocal analysis \cite{Frikel2013} the accurately reconstructed edges are those along which the X-rays are tangent, while blurring occurs across edges that have no tangent X-rays. 
\end{enumerate}

\subsection{Step 1: Total Variation regularization}
\label{subsec:Reconstruction_TV}
\begin{figure}[tb]
    \centering
    \includegraphics[width=0.9\textwidth]{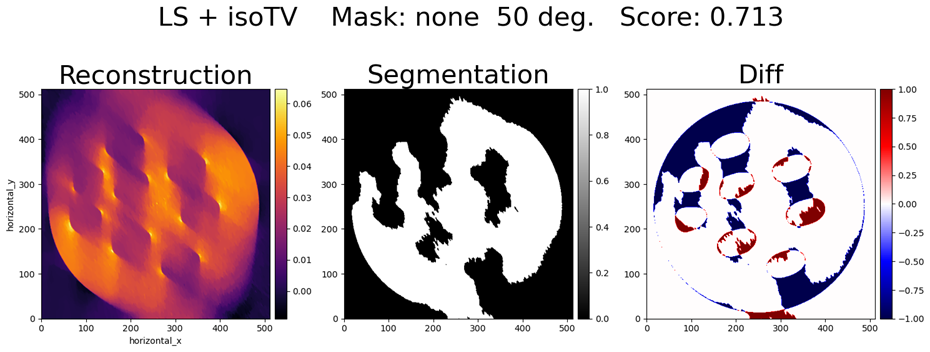}
    \caption{Reconstruction with isotropic TV of sample A with 50$^\circ$ data. Left: reconstruction. Center: segmentation. Right: Difference between segmentation and ground truth segmentation.}
    \label{fig:isotropic_tv}
\end{figure}

We start by seeking to make use of \ref{list:pk-homogeneous} by employing 
TV regularization (\cref{sec:TV}) to favor an image with homogeneous regions with sharp boundaries. We formulate the problem
\begin{align}
    \argmin_{\im} \qquad \omega\, \|\sysmat\im-\sino\|_2^2 + \alpha \, \mathrm{TV}(\im).
    \label{eqn:TV}
\end{align}
Details on implementation with CIL are given in \cref{sec:implementation}. The result (\cref{fig:isotropic_tv}) shows a substantial improvement (score 0.713) compared to 0.430 of FDK (\cref{fig:fdk_score}). However, TV fails to dramatically reduce the directional blurring and in the segmentation some pixels outside the disk toward the lower boundary are misclassified as well as many inside.

\subsection{Step 2: Constraining the reconstruction to a disc}\label{sec:constraint}

 We note a large impact on the score from points around and outside the border of the disk. 
According to \ref{list:pk-disc} we should always have zero-valued pixels outside. To prevent pixels outside from being classified as acrylic we sought to include constraints on pixel values. All pixel values should be greater than or equal to zero. Pixels that are clearly outside could also have an upper bound of 0 enforced (effectively forcing them to zero). Pixels that may be inside can have an upper bound of the linear attenuation coefficient $\mua = 0.0409$ mm$^{-1}$ found for acrylic enforced. 

We initially used a simple disk centered in the center of the reconstruction image with a diameter of 97\% of the image width (about 75 mm) as all training data disks were contained within this. We represented this using the mask $\maskvec$ - a binary mask image with values 1 representing interior pixels and 0 outside pixels. We incorporated this into the optimization problem as zero lower bound and mask-based upper bound disk constraint:
\begin{subequations}
    \begin{alignat}{1}
    \argmin_{\im} \qquad & \omega \,\|\sysmat \im-\sino\|_2^2 + \alpha\, \mathrm{TV}(\im)  \\
    \text{subject to} \qquad &
    \zerovec \leq \im \leq \mua\maskvec.
\end{alignat}
\end{subequations}

While the 97\% disk made some improvement scoring 0.868 on sample A with 50$^\circ$ data, a significant number of pixels at the boundary of the disk were not correctly reconstructed and segmented, mainly because the sample is not guaranteed to be centered and the size of the mask disk is larger than the actual acrylic disk. Therefore, we developed the following procedure based on \cite{circle_fit} to estimate the actual center and radius of the disk, hoping to achieve a more tightly fitting disk.

\begin{figure}[tb]
    \centering
    \includegraphics[width=0.9\textwidth]{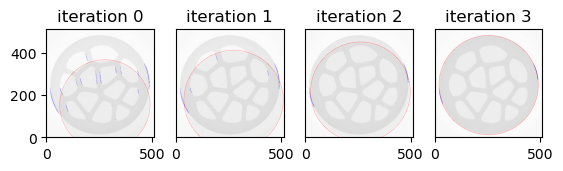}
    \caption{Fitting of the circle with the iterative method described in \cref{sec:constraint}, for the training sample B. In this example 4 iterations are required. In each plot we display: the fitted circumference in red; the edges used by the algorithm in the current step to calculate the circle center and radius, in blue; the reconstruction of the complete data set;  the circle, whose radius is 4 pixels less than the fitted one, which determines which of the edges will be removed at the next iteration.  }
    \label{fig:circle_fit}
\end{figure}

The limited angle FDK reconstruction of the data provided by the organizers shows that roughly parallel  to the main direction of propagation of the X-Rays there are sharp edges. We used these edges to estimate the center and radius of the disk for any data set being reconstructed, with the following algorithm:

\begin{enumerate}
    \item Estimate the location of the sharp edges: (a) Reconstruct the data provided with FDK;
    (b) calculate the magnitude of the gradient of the reconstruction;
    (c) threshold with the Otsu algorithm the magnitude of the gradient of the reconstruction to extract the edge points.
    \item Fit center and radius of a disk to the edge points using \cite{circle_fit}. To discard the edge points due to the internal holes, see \cref{fig:circle_fit} we use the following iterative process: 
    \item Remove from the edge points the ones that are within a circle with radius smaller by 4 pixels from the one found at previous step. 
       Repeat until the number of edge points used to fit the circle does not change.
\end{enumerate}

\begin{figure}[tb]
    \centering\includegraphics[width=0.9\textwidth]{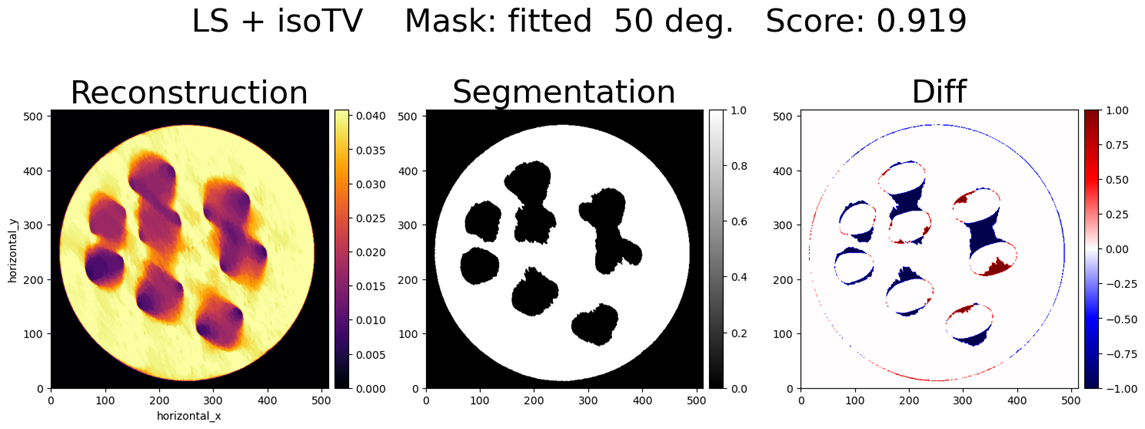}
    \caption{Reconstruction with isotropic TV and disk constraint of sample A with 50$^\circ$ data. Left: reconstruction. Center: segmentation. Right: Difference between segmentation and ground truth segmentation.}
\label{fig:isotropic_tv_disc}
\end{figure}

The result of the reconstruction with the fitted disk can be seen in figure \ref{fig:isotropic_tv_disc}; the score has gone up dramatically to 0.919. Not only are pixels outside now correctly classified but so are many more pixels inside. The improvement achieved by using the fitted disk over the $97\%$ one can be seen by comparing \ref{IsoTV}, which uses the fitted, and \ref{IsoTVconservative} in \cref{fig:challenge-results-table}.

This circle-fitting algorithm removes parts of the external edges that should be used to fit the center and radius of the circle, and a slight effect of this is noticeable with a residual error around the edge in the difference of the ground truth and our segmentations, \cref{fig:isotropic_tv_disc} rightmost column. We considered the algorithm to be sufficiently robust for the challenge, but it could be enhanced by a better rejection of the internal edge data logic.

\subsection{Step 3: Single-directional TV}
While quality has improved, the directional blur still remains.
To further improve the scoring of our algorithm, we sought to exploit \ref{list:pk-directional} by enforcing additional edge-preservation specifically in the orthogonal direction with respect to the central ray in the X-ray fan in the middle of the angular range for the limited angle dataset. We did this using single-directional anisotropic TV regularisation. 

\begin{figure}[tb]
    \centering
    \includegraphics[width=0.9\textwidth]{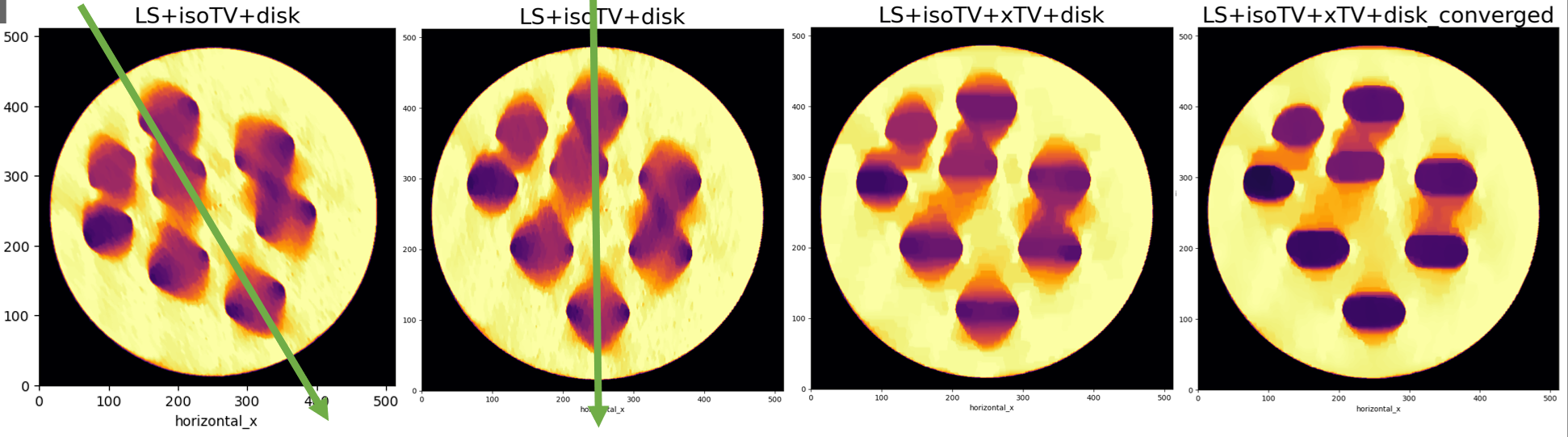}
    \caption{From left to right: isotropic TV reconstruction with, as green arrow the direction of the center X-ray beam in the fan in the middle of the angular range for the limited angle dataset (i.e. the central direction); rotated isotropic such that the y axis is parallel to the central direction (in green); isotropic plus anisotropic TV reconstruction as submitted to the challenge; isotropic plus anisotropic TV reconstruction at convergence, see \ref{subsec:par_det_and_convergence}. 
    \label{fig:anisotropic_tv_method}
    }
\end{figure}

The central direction was not necessarily perpendicular to any of the image axes and CIL currently does not allow directional TV in an arbitrary direction. To apply the anisotropic regularisation we aligned the $y$ axis of the reconstructed image with the propagation direction of the X-rays, by offsetting the rotation angles. 
The anisotropic regularisation could be applied along the $x$ axis of the rotated image. The reconstruction obtained with this procedure was then rotated back to calculate the score, see fig \ref{fig:anisotropic_tv_method}.

\begin{figure}[tb]
    \centering\includegraphics[width=0.9\textwidth]{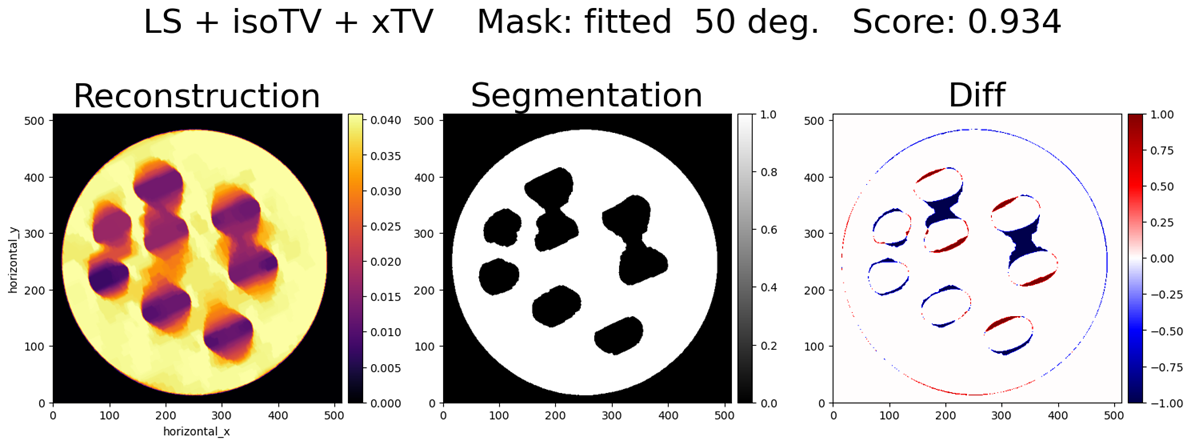}\\\includegraphics[width=0.9\textwidth]{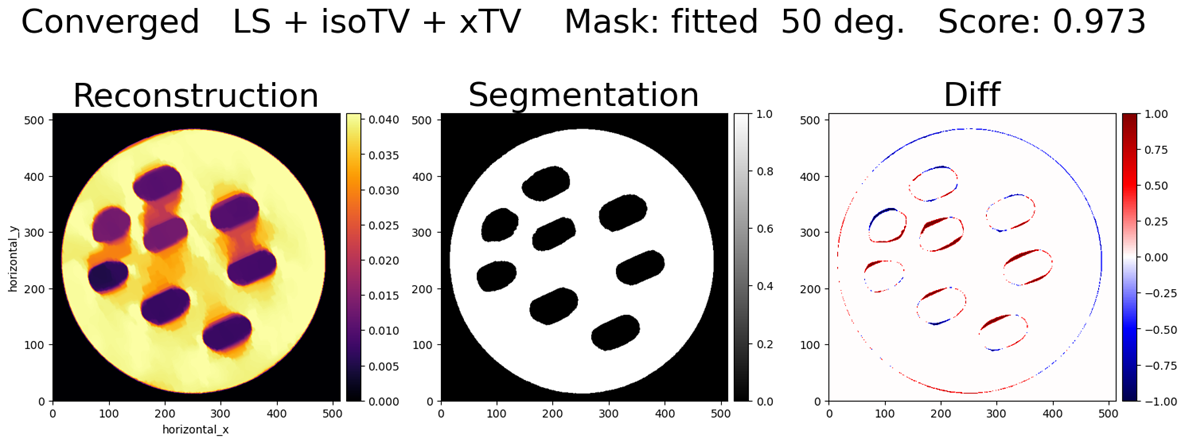}
    \caption{Reconstruction with isotropic and  anisotropic TV with disk constraint of sample A with 50$^\circ$ data. The first row presents the results of the anisotropic TV algorithm as submitted to the challenge.
    The second row presents the results of the same algorithm at full convergence of the PHDG algorithm, resulting in an improvement of the score of about 4\% with respect to the submitted algorithm, see section \ref{subsec:par_det_and_convergence}.}
    \label{fig:anisotropic_tv}
\end{figure}

The optimisation problem solved in the case of isotropic and anisotropic TV with disk constraint is the following, where $\mathrm{TV}_x$ denotes the single-sided anisotropic TV where finite differences are only taken in the $x$-direction: 
\begin{subequations}
\label{eq:singleTV}
\begin{alignat}{1}
\argmin_{\im} \qquad & \omega\, \|\sysmat\im-\sino\|_2^2 + \alpha\, \mathrm{TV}(\im) + \beta \,\mathrm{TV}_x(\im) \label{eq:singleTVopti} \\
     \text{subject to}  \qquad  &  \zerovec \leq \im \leq  \mua \maskvec \label{eq:singleTVcon} 
\end{alignat}
\end{subequations}
The results can be seen in the top row of \cref{fig:anisotropic_tv}; the score has gone further up to 0.934. This is the result of the final submitted algorithm.
After submission, we noticed that for some phantoms the submitted algorithm unfortunately had not fully converged. As explained in \cref{subsec:par_det_and_convergence}, we were able to obtain the results in the bottom row of \cref{fig:anisotropic_tv} for which the score further increased to 0.973.

\subsection{Implementation in CIL}\label{sec:implementation}

In this section, we describe how to solve the specified optimization problems of the previous sections. We focus on the final one \cref{eq:singleTV}; the rest can be solved in the same way, by omitting the relevant terms. 

CIL provides a range of optimization algorithms ranging from gradient descent over FISTA (Fast Iterative Shrinkage Thresholding) \cite{BeckTeboulle} and the Primal-Dual Hybrid Gradient (PDHG) \cite{ChambollePock} method, among others. Here, we employ PDHG as it offers great flexibility to solve all the combinations of different data fidelities, regularizers and constraints that we considered for the challenge.

We recall that PDHG solves the following optimization problem:
\begin{equation}
\argmin_{\im} \qquad F(\bm{K}\im) + G(\im),
\end{equation}
where $F$ and $G$ are convex, possibly nonsmooth, functionals with simple proximal operators and $\bm{K}$ is a continuous and linear operator. To use PDHG as implemented in CIL, we need to provide the ($F$, $\bm{K}$, $G$). A very useful property is that PDHG allows the expansion of the first term as a separable sum of an arbitrary number of terms with a simple proximal:
\begin{align}
    F(\bm{K}\im) = \sum_j F_j(\bm{K_j}\im).
\end{align}
To solve \cref{eq:singleTV} we recast on the form required by PDHG, first including the constraints into the objective function
\begin{align}
\argmin_{\im} \qquad & \omega \,\|\sysmat\im-\sino\|_2^2 + \alpha\, \|\dmat \im\|_{2,1} + \beta \,\|\dmat_x \im\|_1 + \indifun{\mathcal{M}}(\im),
     \label{eq:PDHGsingleTV}
\end{align}
where $I_\mathcal{M}$ is the convex indicator function
\begin{align}
    \indifun{\mathcal{M}}(\im) = 
  \begin{cases}
    0 & \text{if}\quad \im \in \mathcal{M}, \\
    +\infty & \text{otherwise},
  \end{cases}
\end{align}
for a closed convex set $\mathcal{M}$, for which the proximal mapping is simply the Euclidean projection. The convex set $\mathcal{M}$ expresses the non-negativity constraint and the mask upper bound:
\begin{align}
    \mathcal{M} = \{~ \im \in \mathbb{R}^{512^2}  ~\vert~ \zerovec \leq \im \leq \mua \maskvec ~ \}.
\end{align}
We can now express the triplet needed for PDHG as follows:
\begingroup
\renewcommand*{\arraystretch}{1.2}
\begin{align}
    F = \begin{pmatrix}
        F_1 \\ 
        F_2 \\ 
        F_3
    \end{pmatrix} = \begin{pmatrix}
        \omega\,\|\cdot-\sino\|^{2} \\ 
        \alpha\,\|\cdot\|_{2,1} \\ 
        \beta\,\|\cdot\|_{1}
        \end{pmatrix}, \qquad \bm{K} = \begin{pmatrix}
            \bm{K_1} \\ 
            \bm{K_2} \\ 
            \bm{K_3} 
        \end{pmatrix} =
        \begin{pmatrix}
            \sysmat \\ 
            \dmat \\ 
            \dmat_x 
        \end{pmatrix}, \qquad G = \indifun{\mathcal{M}}(\cdot).
\end{align}
\endgroup
The CIL code below sets up this problem, where we emphasize how the code syntax matches the mathematical expressions closely. After specifying the triplet we choose primal and dual steps sizes $\sigma$ and $\tau$ and set up the PDHG algorithm with a zero initial image and run 2000 iterations:
\begin{center}
\begin{tcolorbox}[
    enhanced,
    attach boxed title to top center={yshift=-2mm},
    colback=darkspringgreen!20,
    colframe=darkspringgreen,
    colbacktitle=darkspringgreen,
    title=PDHG: \eqref{Iso1DAnisoTV},
    text width = 12.7cm,
    fonttitle=\bfseries\color{white},
    boxed title style={size=small,colframe=darkspringgreen,sharp corners},
    sharp corners,
]
\begin{minted}{python}

# Separable sum as BlockFunction
F1 = omega*L2NormSquared(b=data)
F2 = alpha*MixedL21Norm()
F3 = beta*L1Norm()
F  = BlockFunction(F1, F2, F3)

# BlockOperator
A  = ProjectionOperator(ig, data.geometry)
D  = GradientOperator(ig)
Dx = FiniteDifferenceOperator(ig, direction='horizontal_x')
K  = BlockOperator(A, D, Dx)

# Constraint
G = IndicatorBoxPixelwise(lower=0.0, upper=muA*m)

# Primal-dual step sizes
sigma = 1.0
tau   = 1.0/(sigma*K.norm()**2)

# Set up and run optimization algorithm
algo = PDHG(initial=ig.allocate(0.0), f=F, g=G, operator=K, 
            sigma=sigma, tau=tau, max_iteration=2000)
algo.run(2000)
\end{minted}
\end{tcolorbox}
\end{center}
The \texttt{ProjectionOperator} $\sysmat$ is GPU accelerated using either the ASTRA \cite{vanAarle2016} or TIGRE \cite{Biguri2016} backend.
Version v22.0.0 of CIL was used in the challenge and no new functionality was added except to support  pixel-wise constraints achieved by the new class  \texttt{IndicatorBoxPixelwise}. This functionality has been integrated into the existing CIL \texttt{IndicatorBox} from version v22.2.0.

\subsection{The submitted CIL algorithms}\label{subsec:all_submissions}

We submitted 5 algorithms for the Helsinki Tomography Challenge. Four of them use the $L^{2}$ norm squared and one of them the $L^{1}$ norm for the fidelity term. For regularisation, we use TV  in both isotropic and anisotropic formulations, and for constraints all employ non-negativity and a disk-shaped mask upper bound. All problems are solved by PDHG. 
We summarize all the algorithms below:
\begin{alignat*}{2}
 \argmin_{\im} \qquad & \omega\,\|\sysmat \im - \sino\|^{2}_{2} + \alpha\,\|\dmat\im\|_{2,1} &&+ \indifun{\mathcal{M}}(\im) 
    \label{IsoTV}\tag{CIL Algo 1}\\
    \argmin_{\im}\qquad & \omega\,\|\sysmat \im - \sino\|^{2}_{2} + \alpha\,\|\dmat\im\|_{2,1} + \beta\,\|\dmat_{x}\im\|_{1} &&+ \indifun{\mathcal{M}}(\im)
    \label{Iso1DAnisoTV}\tag{CIL Algo 2}\\
 \argmin_{\im} \qquad & \omega\,\|\sysmat \im- \sino\|_{1} + \alpha\,\|\dmat \im\|_{2,1} + \beta\,\|\im\|_{1} &&+  \indifun{\mathcal{M}}(\im)
\label{L1TVL1} \tag{CIL Algo 3}\\
 \argmin_{\im} \qquad & \omega\,\|\sysmat \im - \sino\|_{2}^{2} + \alpha\,\|\dmat \im\|_{1,1} - \beta\,\|\dmat \im\|_{2,1} &&+  \indifun{\mathcal{M}}(\im)\label{wdIsoAniso}\tag{CIL Algo 4}\\
  \argmin_{\im} \qquad & \omega\,\|\sysmat \im - \sino\|^{2}_{2} + \alpha\,\|\dmat\im\|_{2,1} &&+ \indifun{\mathcal{M}^{97}}(\im) 
    \label{IsoTVconservative}\tag{CIL Algo 5}
\end{alignat*}
\ref{IsoTV} employs just TV-regularization and the disk-constraint while \ref{Iso1DAnisoTV} adds the single-directional TV previously presented. In \ref{L1TVL1} we experimented with the $L^1$ norm data fidelity in combination with both TV and $L^1$ regularization to encourage both sharp edges and sparsity. \ref{wdIsoAniso} employs the difference-based anisotropic-isotropic TV combination described at the end of \cref{sec:TV}. Finally, \ref{IsoTVconservative} is identical to \ref{IsoTV} except the mask $\mathcal{M}^{97}$ used in the upper bound constraint is the disk having a diameter of 97\% of the image width. This was to provide a more conservative fallback algorithm in case the disk-fitting method would fail for the limited-angle evaluation data.

In general, on the training data set, the performance of the winning algorithm was comparable with the other algorithms submitted by the team, while it always outperformed them on the evaluation datasets.

\subsection{Algorithm testing and determination of parameters}
\label{subsec:par_det_and_convergence}

Each reconstruction algorithm contains a number of parameters in the optimization problem to be tuned for optimal quality. We note that the setup of the challenge with evaluation on unseen data with seven different levels of limited-angle data required fully automated parameter selection. 

Our best performing algorithm \ref{Iso1DAnisoTV} includes three parameters $\omega$, $\alpha$ and $\beta$. In principle one of these is redundant but we set $\omega$ to 90 divided by the angular range in degrees, to effectively normalize the data fidelity, because this term otherwise grows linearly with the number of projections.

For $\alpha$ and $\beta$ we employed a brute-force extensive parameter sweep based on maximizing the score metric across all the four training data sets with limited-angle data simulated at all 7 difficulty levels. This resulted in the constant choice of values to be used at all levels of $\alpha = 0.01$ and $\beta = 0.03$. The same procedure was used for all 5 algorithms submitted by the team.

In addition, the PDHG algorithm requires specification of three internal algorithm parameters, namely the primal and dual step sizes and the number of iterations to run. 
Based on experiments during the challenge, our choice was  step sizes $\sigma = 1.0$ and $\tau = 1/||\bm{A}||_2^2$ and 2000 iterations, which was considered sufficient for convergence. After the submission to the challenge, we tried 2000 iterations with the default step sizes of PDHG defined in CIL, i.e. $\sigma = \tau = 1/||\bm{A}||_2$. The results are in \cref{fig:anisotropic_tv}, where it can be seen that the edges are better determined and the score is slightly further improved. 
However, only on some datasets these alternative step sizes produced an improved result.
\section{Results}
\label{sec:results}

After optimizing the algorithms developed for the four training datasets provided, the algorithms were submitted to the challenge and evaluated by the challenge organizers on a collection of evaluation datasets unseen by participants. 
The increase in difficulty, from level 1 to 7 was achieved by  reducing the angular range from 90$^\circ$ to 30$^\circ$, by increasing the number and shape-complexity of holes as well as by changing their orientation relative to the measured angles. 

\begin{figure}[tb]
\newcommand{\iwidth}{0.15\linewidth}
    \begin{picture}(200,30)(0,0)
    \put(65,17){Case A}
    \put(215,17){Case B}
    \put(365,17){Case C}
    \put(15,2){Ground truth}
    \put(100,2){Result}
    \put(165,2){Ground truth}
    \put(250,2){Result}
    \put(315,2){Ground truth}
    \put(400,2){Result}
    \end{picture}\\    
    \rotatebox{90}{\phantom{00}90 degrees}
    \includegraphics[width=\iwidth]{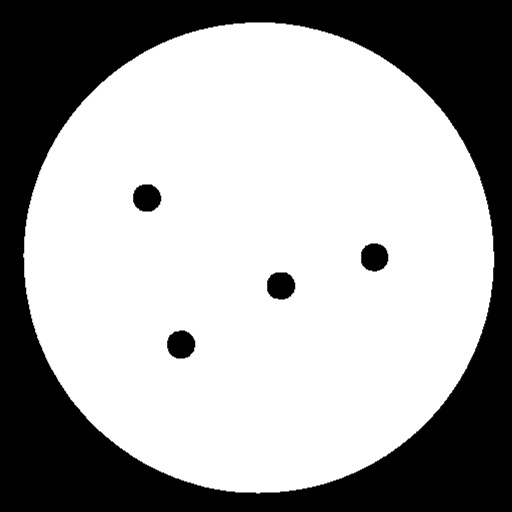} \includegraphics[width=\iwidth]{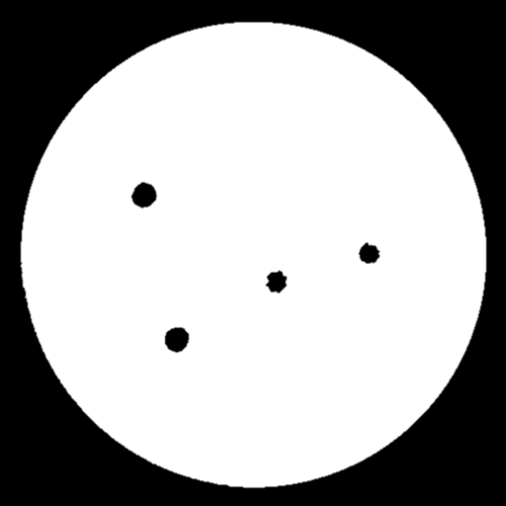} \quad
    \includegraphics[width=\iwidth]{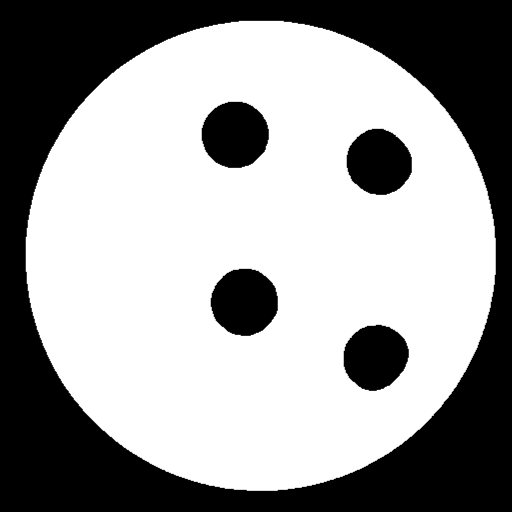} \includegraphics[width=\iwidth]{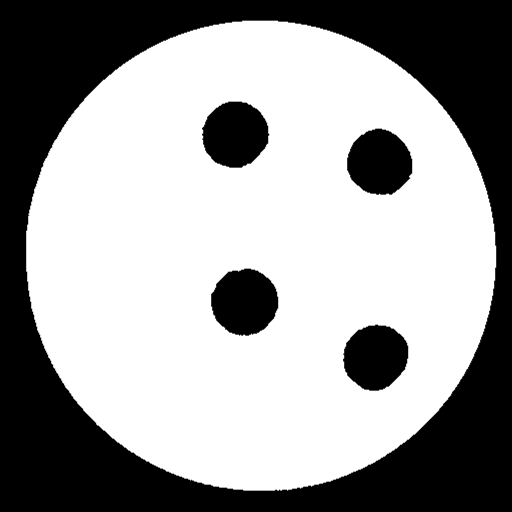}
    \quad 
    \includegraphics[width=\iwidth]{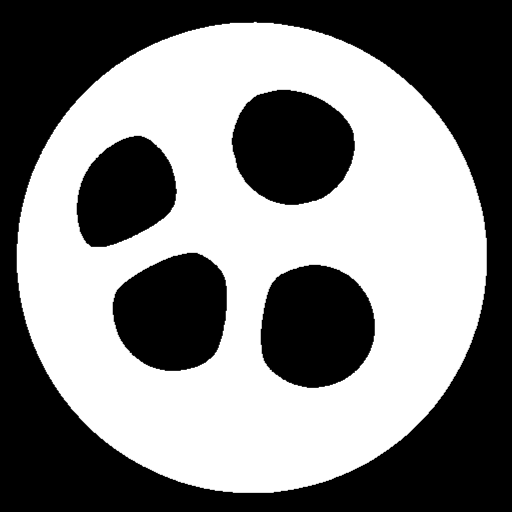} \includegraphics[width=\iwidth]{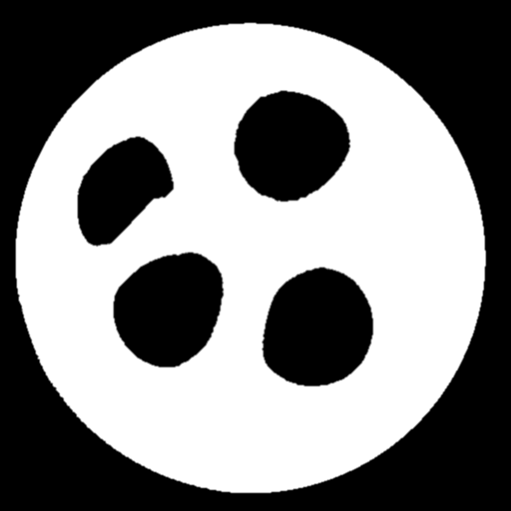}\\[0.03cm] 
    \rotatebox{90}{\phantom{00}80 degrees} 
    \includegraphics[width=\iwidth]{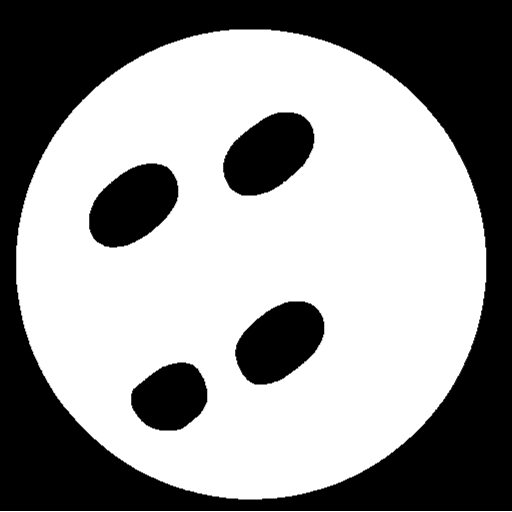} \includegraphics[width=\iwidth]{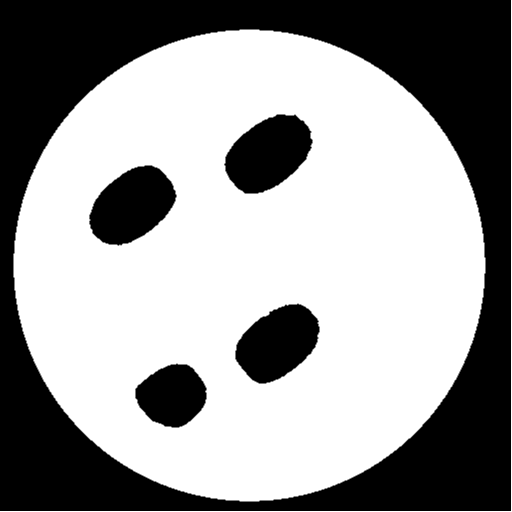} 
    \quad
    \includegraphics[width=\iwidth]{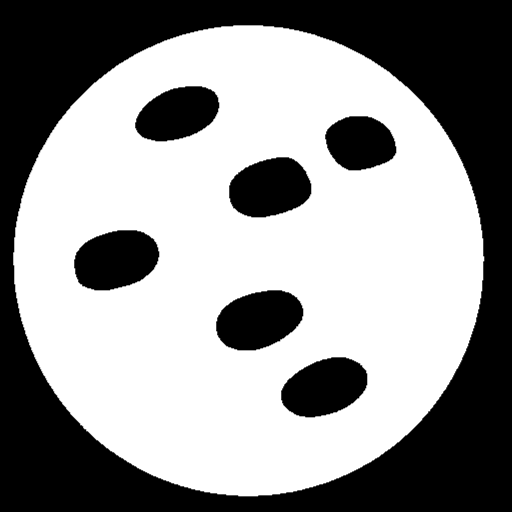} 
    \includegraphics[width=\iwidth]{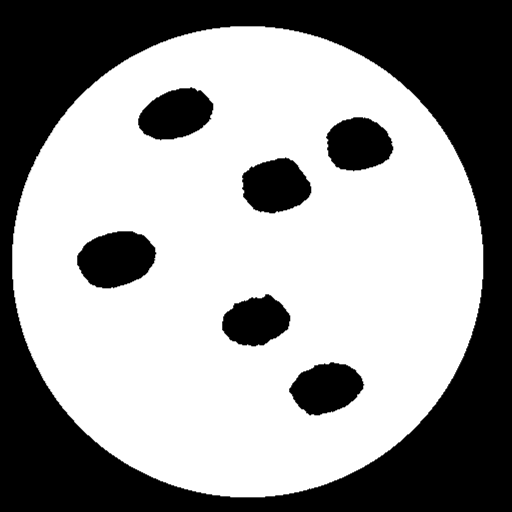} 
    \quad
    \includegraphics[width=\iwidth]{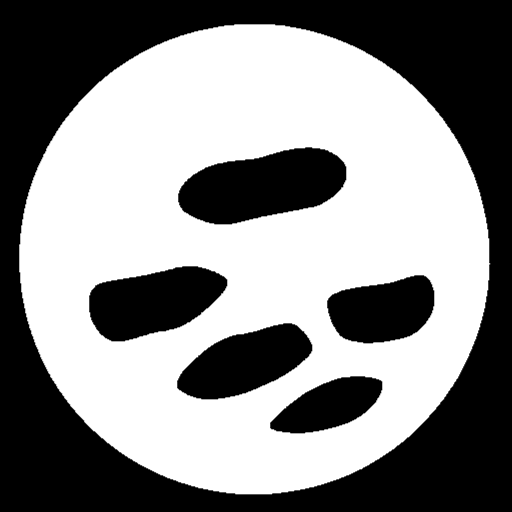} 
    \includegraphics[width=\iwidth]{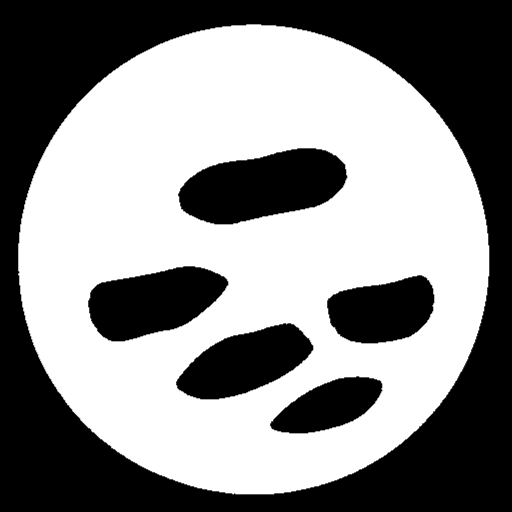}\\[0.03cm]
    \rotatebox{90}{\phantom{00}70 degrees}
    \includegraphics[width=\iwidth]{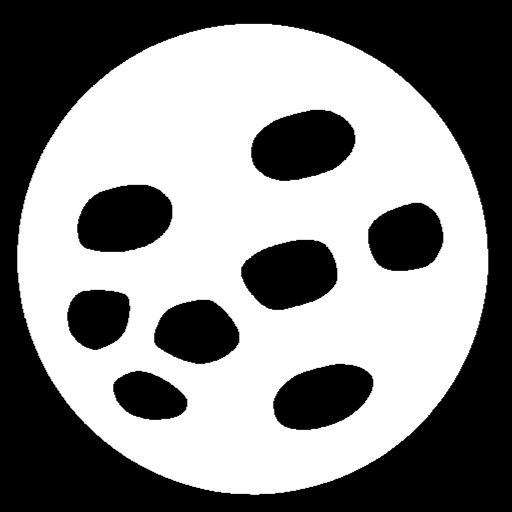} 
    \includegraphics[width=\iwidth]{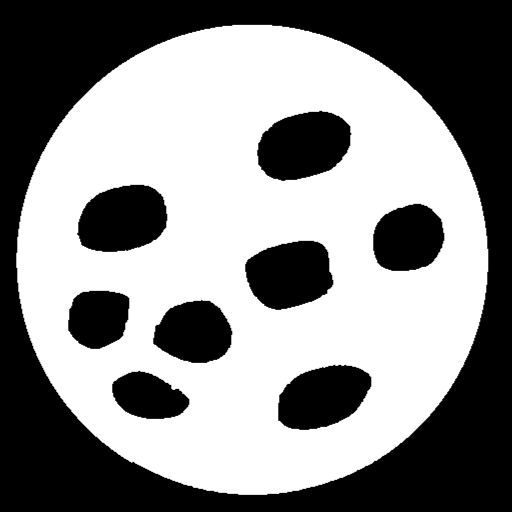}
    \quad
    \includegraphics[width=\iwidth]{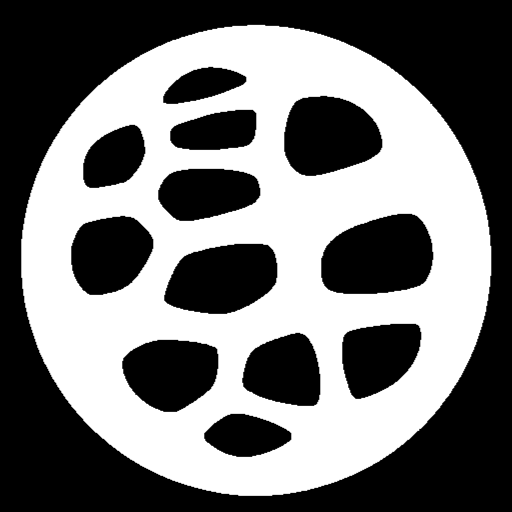}
    \includegraphics[width=\iwidth]{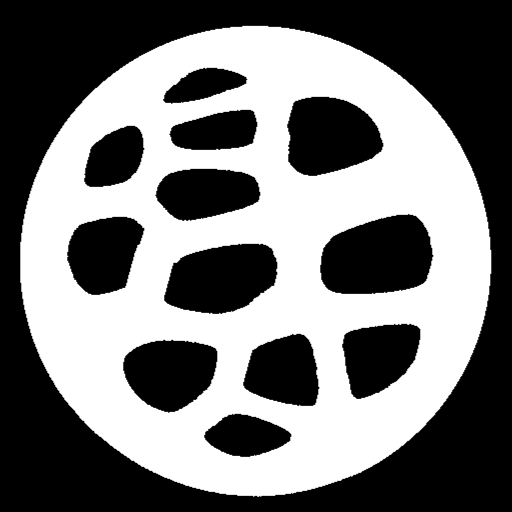} 
    \quad
    \includegraphics[width=\iwidth]{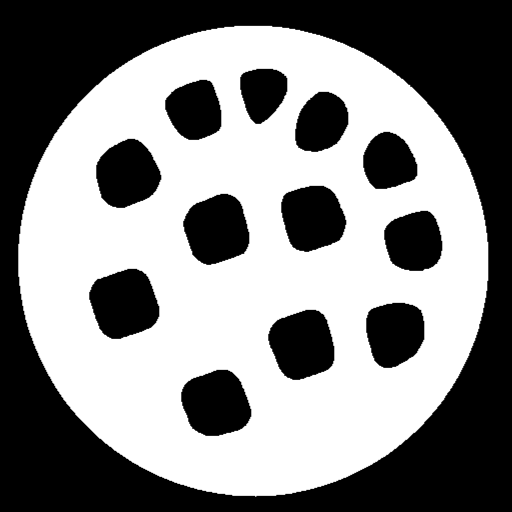} \includegraphics[width=\iwidth]{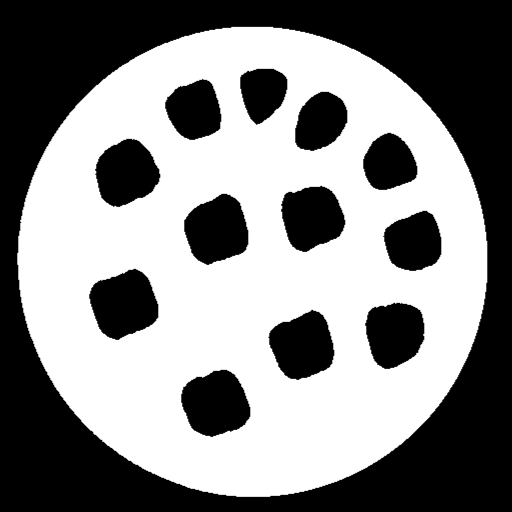} \\[0.03cm]
    \rotatebox{90}{\phantom{00}60 degrees}
    \includegraphics[width=\iwidth]{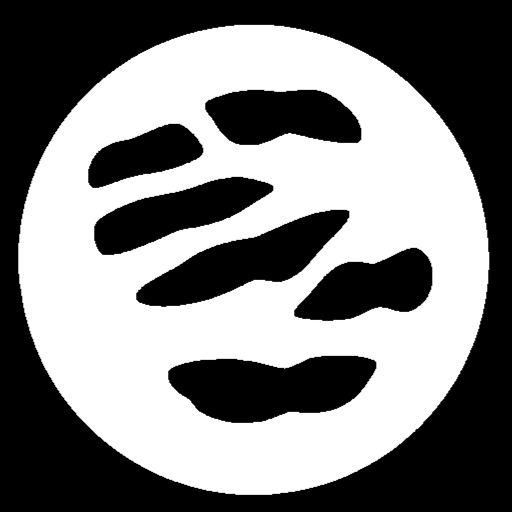} 
    \includegraphics[width=\iwidth]{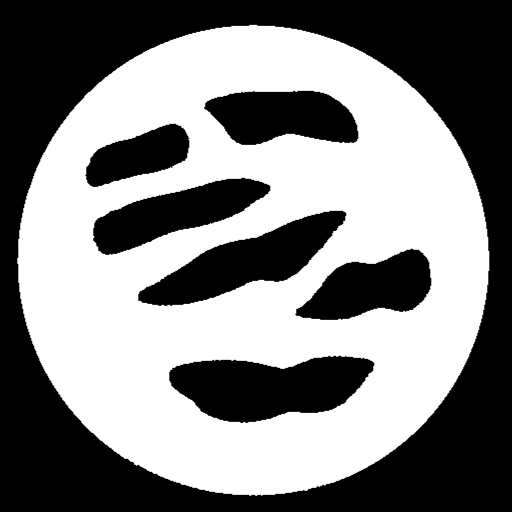}
    \quad
    \includegraphics[width=\iwidth]{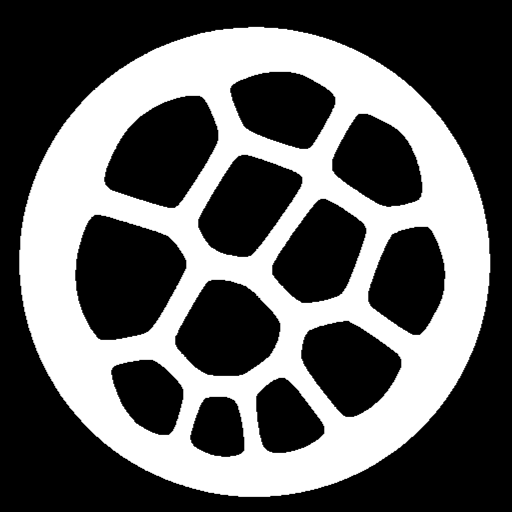} 
    \includegraphics[width=\iwidth]{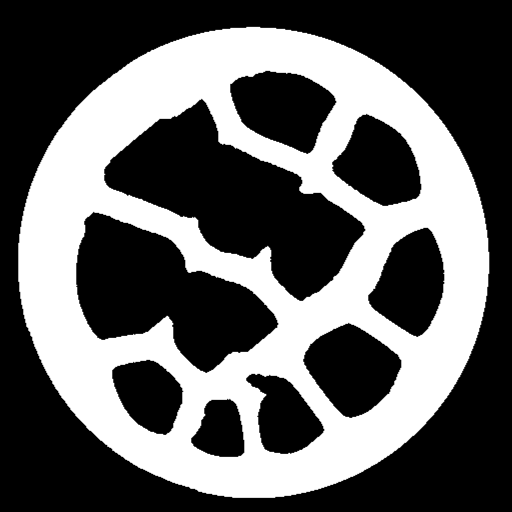}
    \quad
    \includegraphics[width=\iwidth]{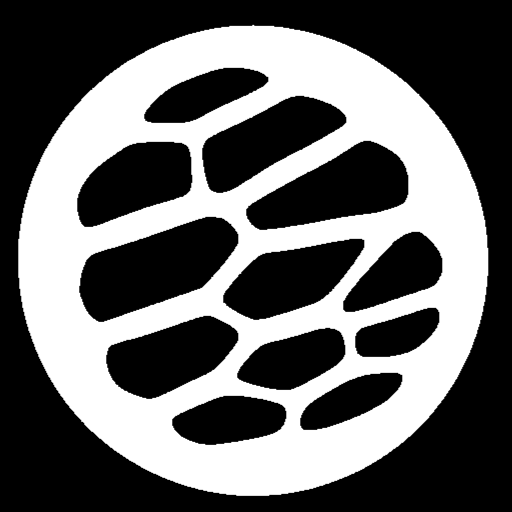} 
    \includegraphics[width=\iwidth]{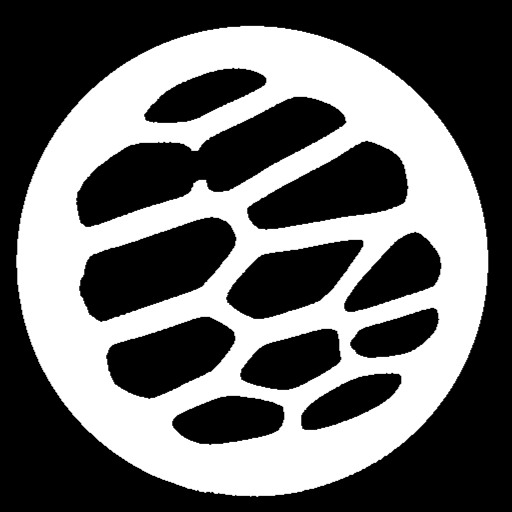}\\[0.03cm]
    \rotatebox{90}{\phantom{00}50 degrees}
    \includegraphics[width=\iwidth]{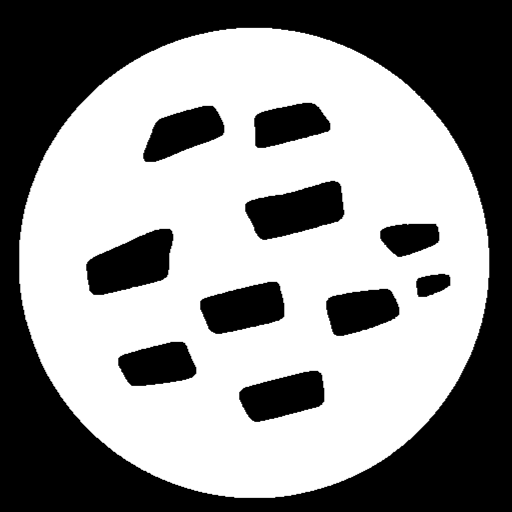} 
    \includegraphics[width=\iwidth]{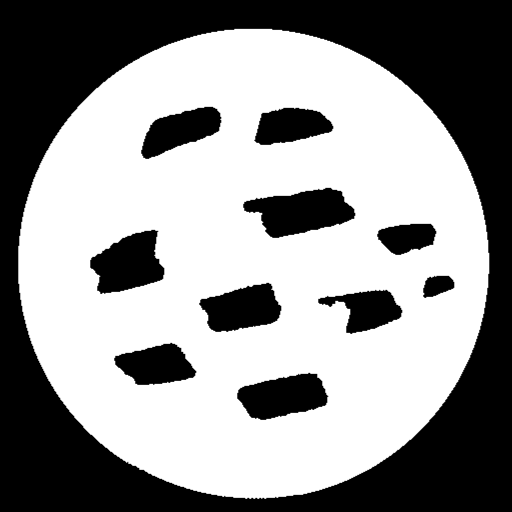}
    \quad
    \includegraphics[width=\iwidth]{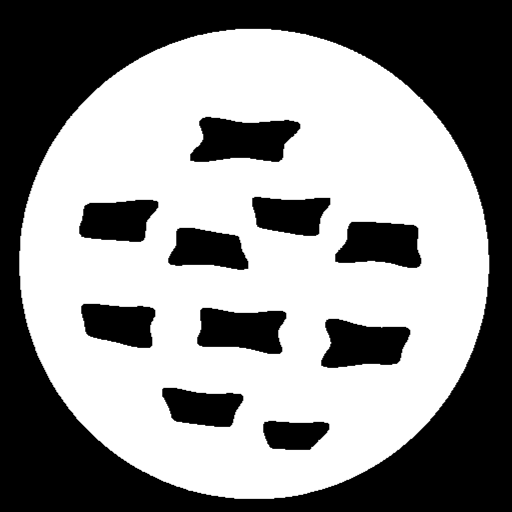} 
    \includegraphics[width=\iwidth]{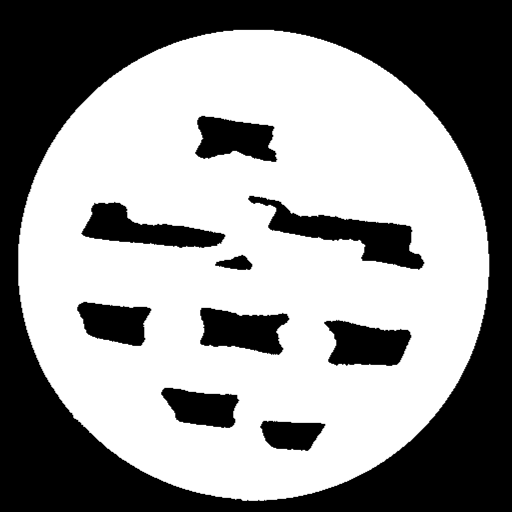}
    \quad
    \includegraphics[width=\iwidth]{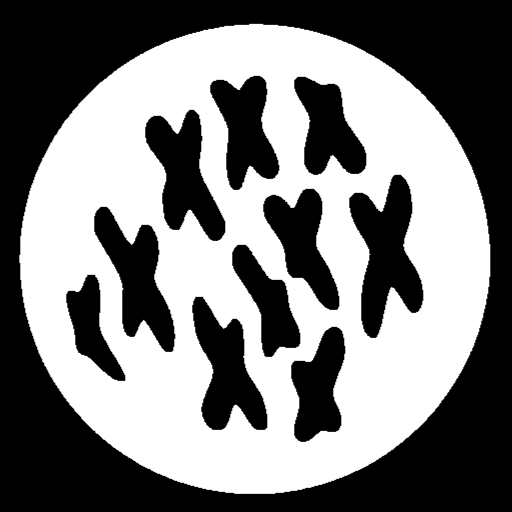} \includegraphics[width=\iwidth]{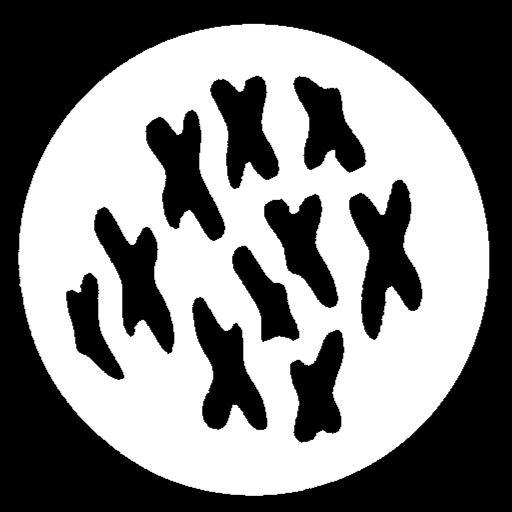}\\[0.03cm]
    \rotatebox{90}{\phantom{00}40 degrees} 
    \includegraphics[width=\iwidth]{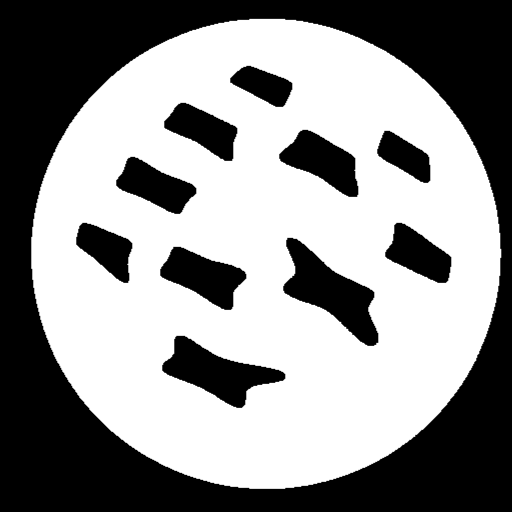} \includegraphics[width=\iwidth]{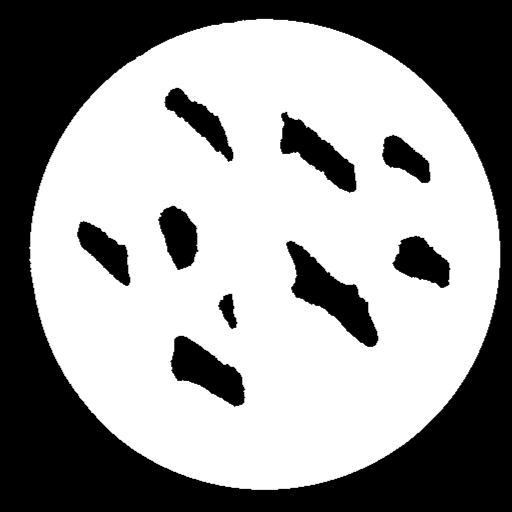}
    \quad
    \includegraphics[width=\iwidth]{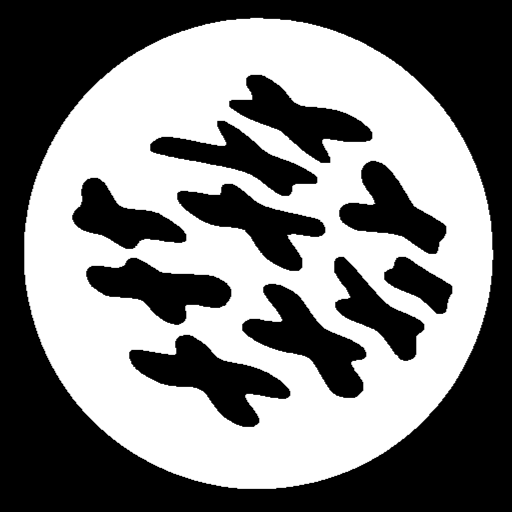} 
    \includegraphics[width=\iwidth]{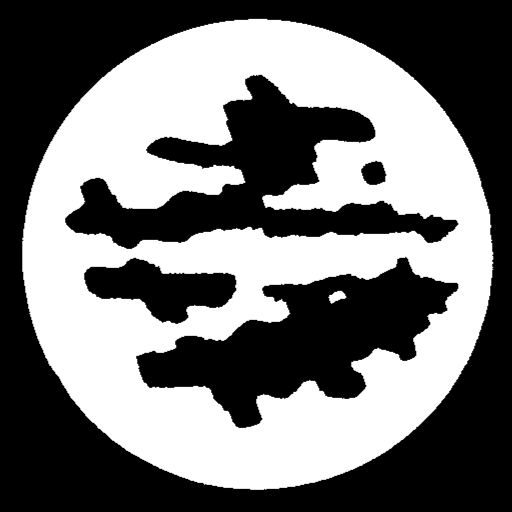}
    \quad
    \includegraphics[width=\iwidth]{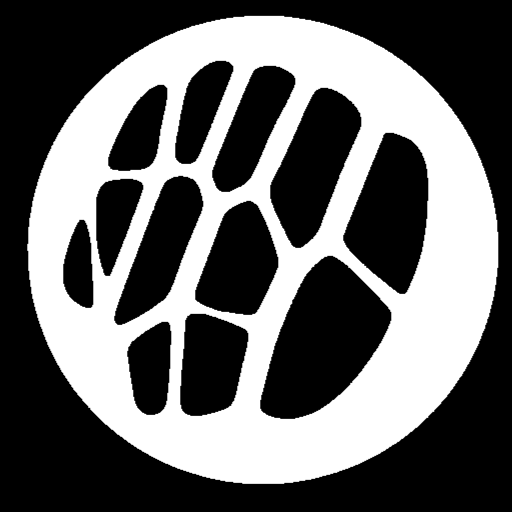} 
    \includegraphics[width=\iwidth]{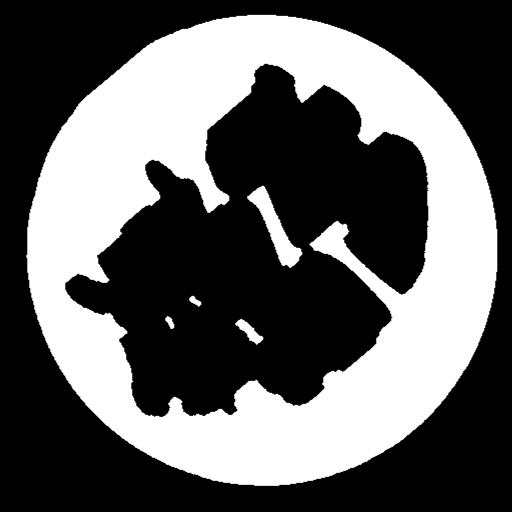}\\[0.03cm]
    \rotatebox{90}{\phantom{00}30 degrees} 
    \includegraphics[width=\iwidth]{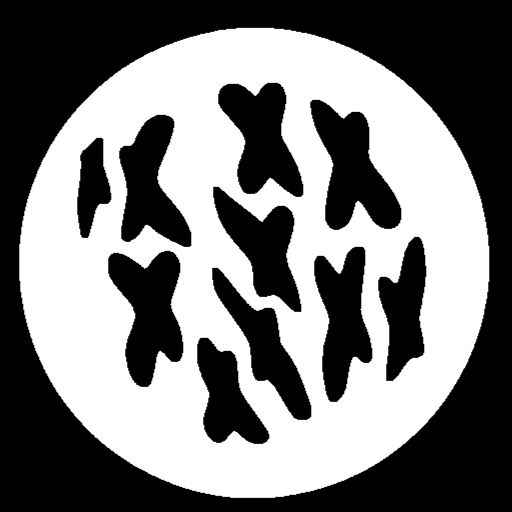} 
    \includegraphics[width=\iwidth]{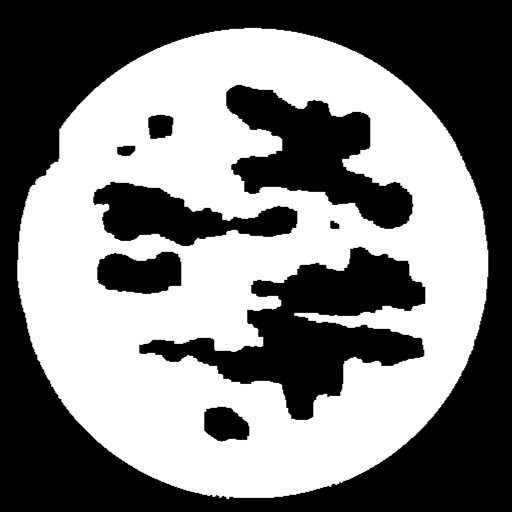}
    \quad
    \includegraphics[width=\iwidth]{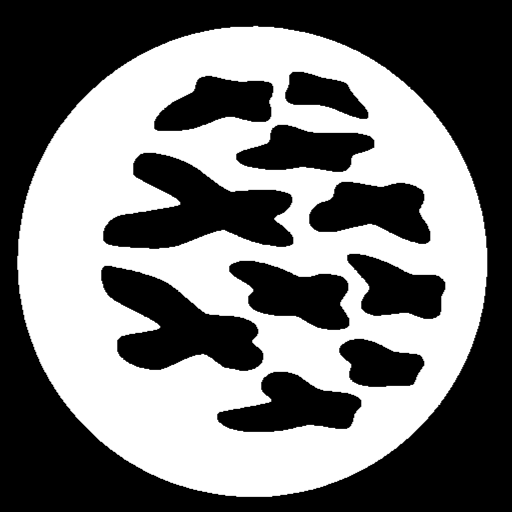} 
    \includegraphics[width=\iwidth]{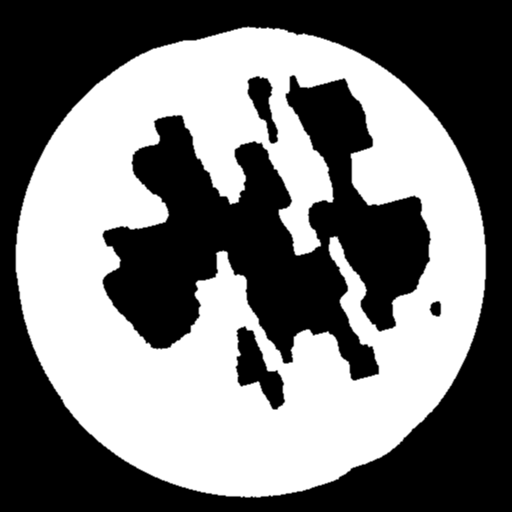}
    \quad
    \includegraphics[width=\iwidth]{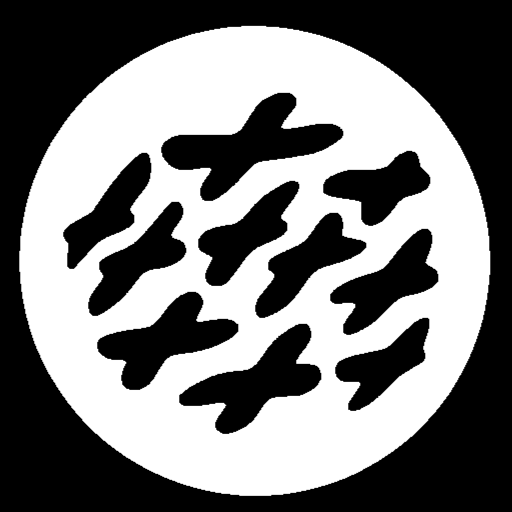} \includegraphics[width=\iwidth]{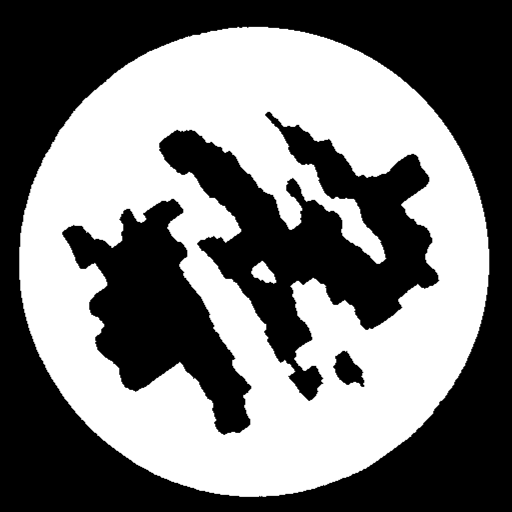}
    \caption{Results of the CIL team's Algorithm 2, compared to ground truth, for evaluation datasets.}
    \label{fig:challenge-results-table}
\end{figure}
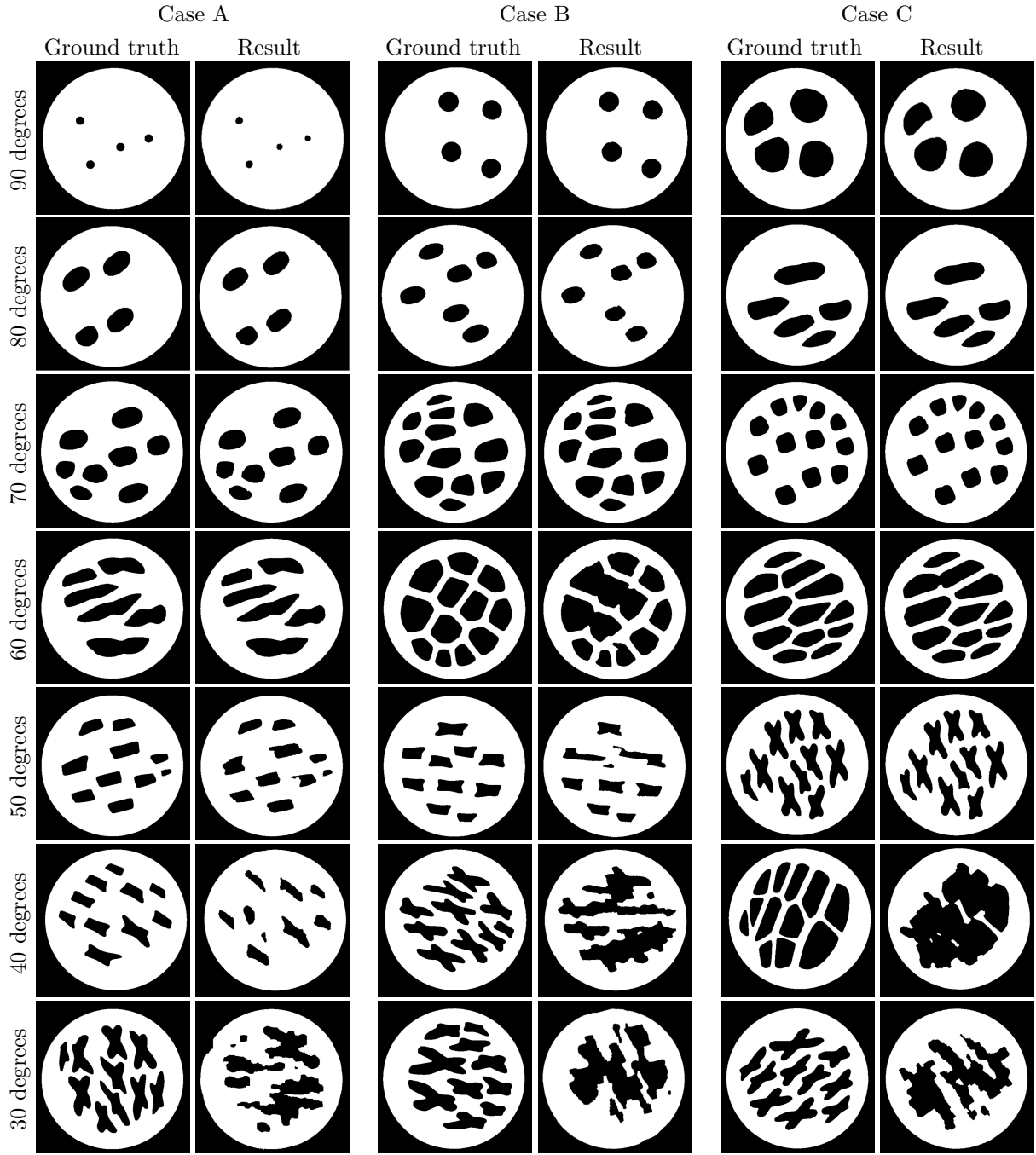

\begin{figure}[tb]
    \centering
    \includegraphics[width=\linewidth]{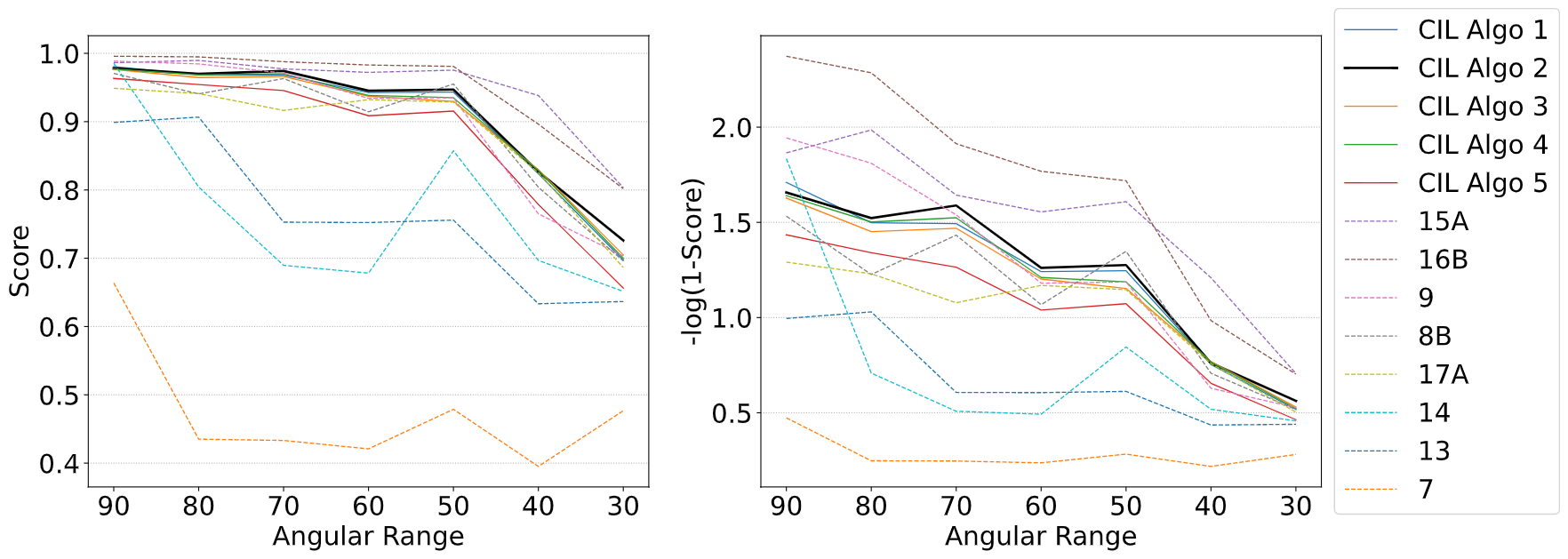}
    \caption{Average scores over the three phantoms at each of the seven levels from $90^\circ$ to $30^\circ$ data for the five CIL algorithms and the highest scoring algorithm from each competing team. Left: Raw score. Right: Scores transformed by $-\log(1-\texttt{score})$ to highlight differences close to 1.  }
    \label{fig:scores_graphs}
\end{figure}

The complete results have been published by the organizers at \url{https://fips.fi/HTC2022.php}. Of the five CIL algorithms, algorithm 2 scored the highest and finished overall third in the competition. The complete set of final segmented images produced by CIL Algorithm 2 along with ground truths for comparison are shown for the evaluation data in \cref{fig:challenge-results-table}. The quantitative scores for all CIL algorithms and the highest scoring algorithm from all participating teams are shown in \cref{fig:scores_graphs}. Segmented images and scores for the training data (for simulated angular ranges $90^\circ$, $60^\circ$ and $30^\circ$) can be found at \url{https://github.com/TomographicImaging/CIL-HTC2022-Algo2}.

\Cref{fig:challenge-results-table} shows near-perfect segmented reconstructions for all three evaluation phantoms at levels $90^\circ$, $80^\circ$ and $70^\circ$ as well as for phantoms A and C at $60^\circ$ and $50^\circ$. For phantom B at $60^\circ$ and $50^\circ$ some of the holes have started merging together and this caused the average score to be reduced. 
At $40^\circ$ phantom A is reconstructed reasonably well, whereas B and C have large holes caused by algorithm not being sufficiently able to handle the strong directional blurring. 
In general, edges to which the limited-angle data contains rays that are tangent are easier to reconstruct and are preserved. Edges that are roughly perpendicular to the central X-ray direction have no tangent rays and tend to be lost in the directional blur. 
At $30^\circ$, the outer shape remains correctly reconstructed, however the voids are poorly reconstructed. At this difficulty level also all other algorithms from competing teams were struggling to produce reasonable reconstructions.

The same trends are reflected in the graph of scores shown in \cref{fig:scores_graphs}. In the left figure, the raw average score is shown for all five CIL algorithms in full line along with the best scoring algorithms from competing teams shown dashed as function of angular range.  The right figure shows the same, except the raw average scores have been transformed by $-\log(1-\cdot)$ in order to emphasize difference between scores close to and just under one. In agreement with the visual inspection, the quantitative score of CIL Algorithm 2 remains almost constant at the first three levels, drops a bit at $60^\circ$ and $50^\circ$ caused by phantom B, before larger drops at $40^\circ$ and $30^\circ$. CIL algorithms 1, 3 and 4 performed similarly but slightly worse, whereas CIL algorithm 5, which was more conservative in that it did not exploit the fitted disk mask constraint but only a general disk of diameter 97\% of the image size, scored yet a bit lower. Similar overall trends can be seen for the competing algorithms, with 1st and 2nd place algorithms scoring noticeably higher at all levels. A notable exception is the algorithm of team 14 which scored very high at the first level but after that dropped to lower scores.

\begin{figure}[tb]
    \centering
    \includegraphics[width=\linewidth]{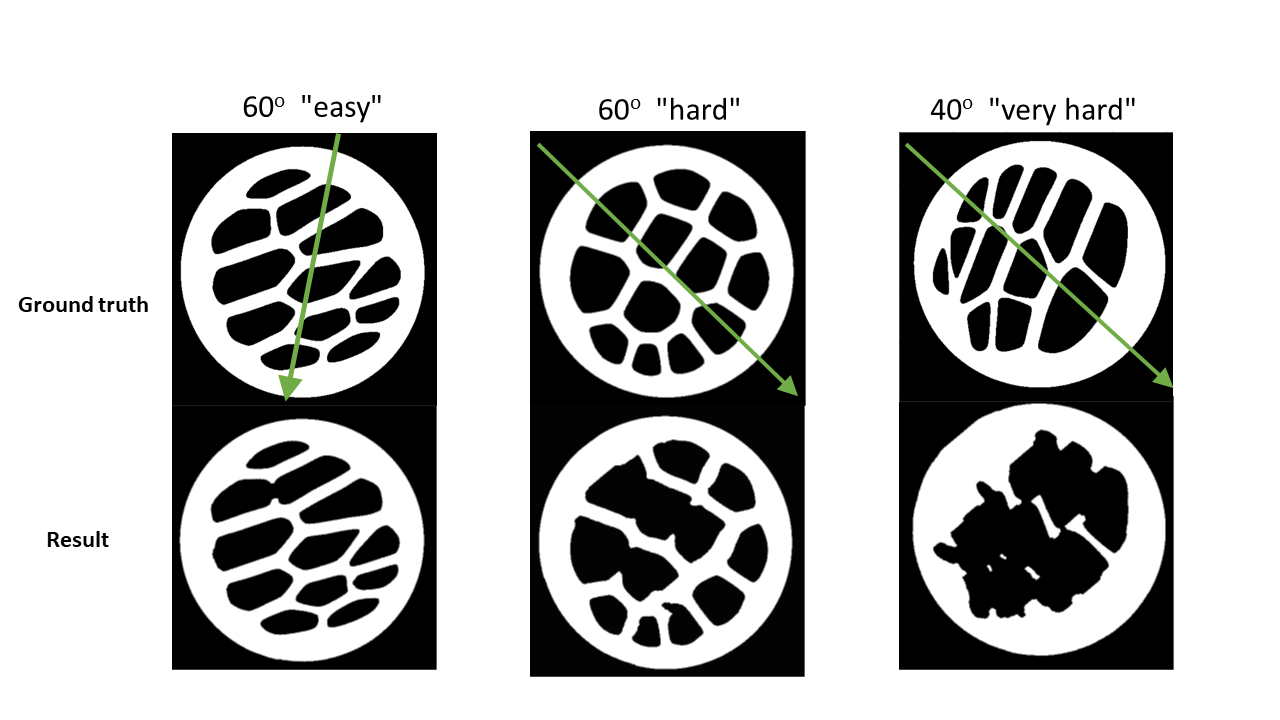}
    \caption{ Each column shows one of the evaluation samples, with ground truth on the top and output of our algorithm on the bottom. Superimposed as green arrow, the central direction of the X-ray cone. Left: level 4 ($60^\circ)$, sample C. Center: level 4 ($60^\circ)$, sample B. Right: level 6 ($40^\circ$), sample C. If most edges are roughly \emph{not} perpendicular to the central direction of the X-ray fan, these are captured by data and score is high (Left: 0.96374). If most edges \emph{are} roughly perpendicular they are harder to reconstruct and score lower (center: 0.92751, right: 0.74646). The right column shows that only the structures parallel to the central direction of the X-ray can be recovered.}
    \label{fig:direction_matters}
\end{figure}

Finally, we note that not only the size of the angular range affects the score but also its orientation relative to the features of the image.  \Cref{fig:direction_matters} shows one well-reconstructed case ($60^\circ$, C), one less well ($60^\circ$, B) and one poorly ($40^\circ$, C) along with the orientation of the central X-ray. 
Edges that are orthogonal to the central X-ray tend to be the poorly reconstructed, in agreement with microlocal analysis \cite{Frikel2013}. In the left case, few such edges exist and hence the reconstruction is good. In the center figure some are orthogonal and this is what causes $60^\circ$, B in \cref{fig:challenge-results-table} to be scoring lower than A and C. If as in the right figure, most edges are orthogonal, then the quality will be poor. 

\section{Discussion}
\label{sec:discussion}

The Helsinki Tomography Challenge 2022 offered a very well-designed set of phantoms, experimental setup and image quality metric and objective, making it straightforward to compare the quality produced by algorithms.
The particular design of the phantoms as homogeneous disks with holes provided a setup for which a wide range of algorithms could be considered seeking to incorporate various aspects of the phantom structure as prior information, such as sharp edges, binarity, disk-shape etc. This was an advantage for a model-based approach in which such information can be codified as regularizers or constraints, compared to filtered back-projection that does not allow it. Similarly, the relative simplicity of the phantoms also provided a clear advantage for deep learning based methods, since large amounts of accurate training data can be generated with relative ease.

More complicated imaging subjects such as of porous media, composite materials or cross sections of the human body often contain much more involved structures, multiple gray levels or even phases with textured/gradually varying attenuation. It would be more challenging to develop both model-based as well as learning-based reconstruction methods for such cases, and the quality obtained on the challenge should not be expected at such severe limited-angle scenarios. However, the designing of an algorithm for such cases would follow the exact same principles of seeking to determine and incorporate as much prior information as possible.

The choice of evaluating quality on a segmented image rather than the reconstruction itself also offered a convenient unambiguous setup, which ties to the common use case of using segmentation to quantify e.g. size or surface areas of components of the image. It does carry the implication that the final score not only reflects the quality of the reconstruction algorithm but also of the segmentation algorithm (as well as pre-processing steps), but this is obviously consistent with practical use cases where one is interested in how well the final image solves the intended imaging task rather than individual steps in a computational pipeline. It also offers the opportunity to directly produce a segmented image, rather than a reconstruction followed by a separate segmentation, as indeed several of the competing learning-based methods implemented successfully.

While working on the challenge we explored a variety of different approaches and optimization problems. It was tremendously useful to employ the optimization problem prototyping capabilities of CIL to quickly specify and solve a range of different optimization problems from modular algorithm building blocks such as linear operators, functionals and algorithm implementations, rather than having to implement and validate dedicated algorithms for each desired optimization problem. 

We considered a number of other reconstruction approaches and mention a few here as possible future work.
Firstly, parameter values including regularization parameters, number of iterations, algorithm step sizes, etc. had to be automatically set, as the challenge evaluation setup required a complete main file to be submitted, preventing any manual hand-tuning of parameters on individual evaluation data sets. Without a doubt, our parameter choices could be further improved, as in fact we learned that in some cases our resulting images were suboptimal due to not having fully converged.

Secondly, all phantoms consisted of a disk with holes cut. In the training data, no holes were directly in contact with the outer boundary, rather at some distance. One could try to exploit this by forcing a lower bound of the acrylic linear attenuation also used for upper bound, on pixel values in an annulus of some thickness closest to the boundary. In fact we implemented and tested this but since the challenge specifications did not rule out that holes could occur touching the boundary, we decided to omit this.

Finally, one could try to make a stronger use of the fact that the phantoms were essentially binary. We experimented with combinations of L1-norm regularizers and did include $\|\im\|_1$ in our Algorithm 3 to seek to force some values to zero. Another idea was to use $\|\im-\mua\|_1$ to seek to force pixel values to the acrylic attenuation value, or a combination of the two. These approaches preserve convexity of the optimization problem, but allow values in the complete range $[0,\mua]$. A stronger option might be to employ a DART-type (Discrete Algebraic Reconstruction Technique) method to enforce a binary result \cite{DART2011}. Yet another option would be to employ Multi-Bang regularization, which is non-convex and seeks to enforce the image taking on only values in a specified discrete set \cite{Holman_2020}. In fact, this approach was implemented at the CIL ``Bring Your Own Data'' User Hackathon at the Isaac Newton Institute for Mathematical Sciences, Cambridge, UK in March 2023 and a resulting demo is available at \url{https://github.com/TomographicImaging/CIL-User-Showcase}.

\section{Conclusion}\label{sec:conclusion}

The Core Imaging Library (CIL) developer team designed, implemented and optimized five reconstruction methods for limited-angle CT data in the Helsinki Tomography Challenge 2022 employing the optimization capabilities of CIL. The methods consisted of pre-processing (data renormalization, beam hardening correction and zero padding), reconstruction with model-based methods employing different combinations of data fidelities, regularizers and constraints, and multi-Otsu segmentation. The best performing algorithm employed single-sided directional total variation regularization combined with isotropic total variation, a zero lower bound and a pixelwise upper bound on pixel values with the linear attenuation coefficient of acrylic inside a fitted disk and zero outside. The algorithm produced very high quality reconstructions and segmentations down to 50 degree data, whereas quality at 40 and especially 30 degree data was reduced, yet good enough to finish 3rd in the challenge. 

Overall the results suggest that with careful data pre-processing and design of regularizers and constraints to incorporate all available prior information, model-based reconstruction can achieve comparable performance to deep-learning methods for limited-angle CT reconstruction.



\bibliographystyle{siam}
\bibliography{references}

\end{document}